\documentclass{emulateapj}

\usepackage{color}
\usepackage{multirow}

\usepackage{natbib}

\shorttitle{Galaxy Assembly Bias on the Red Sequence}
\shortauthors{Cooper et al.}

\begin{document}


\title{Galaxy Assembly Bias on the Red Sequence}


\author{
Michael C.\ Cooper\altaffilmark{1,2},
Anna Gallazzi\altaffilmark{3},
Jeffrey A.\ Newman\altaffilmark{4}, 
Renbin Yan\altaffilmark{5}
}

\altaffiltext{1}{Steward Observatory, University of Arizona, 
933 N.\ Cherry Avenue, Tucson, AZ 85721 USA; 
cooper@as.arizona.edu}

\altaffiltext{2}{Spitzer Fellow}

\altaffiltext{3}{Max--Planck--Institut f\"{u}r Astronomie,
  K\"{o}nigstuhl 17, D--69177 Heidelberg, Germany; gallazzi@mpia.de}

\altaffiltext{4}{Department of Physics and Astronomy, University of
  Pittsburgh, 401--C Allen Hall, 3941 O'Hara Street, Pittsburgh, PA 15260
  USA; janewman@pitt.edu}

\altaffiltext{5}{Department of Astronomy and Astrophysics, University
  of Toronto, 50 St.\ George Street, Toronto, ON M5S 3H4, Canada;
  yan@astro.utoronto.ca}

\begin{abstract}

  Using samples drawn from the Sloan Digital Sky Survey, we study the
  relationship between local galaxy density and the properties of
  galaxies on the red sequence. After removing the mean dependence of
  average overdensity (or ``environment'') on color and luminosity, we
  find that there remains a strong residual trend between
  luminosity--weighted mean stellar age and environment, such that
  galaxies with older stellar populations favor regions of higher
  overdensity relative to galaxies of like color and luminosity (and
  hence of like stellar mass). Even when excluding galaxies with
  recent star--formation activity (i.e., younger mean stellar ages)
  from the sample, we still find a highly significant correlation
  between stellar age and environment at fixed stellar mass. This
  residual age--density relation provides direct evidence for an
  assembly bias on the red sequence such that galaxies in
  higher--density regions formed earlier than galaxies of similar mass
  in lower--density environments. We discuss these results in the
  context of the age--metallicity degeneracy and in comparison to
  previous studies at low and intermediate redshift. Finally, we
  consider the potential role of assembly bias in explaining recent
  results regarding the evolution of post--starburst (or
  post--quenching) galaxies and the environmental dependence of the
  type Ia supernova rate.

\end{abstract}


\section{Introduction}
\label{sec_intro}

Over the past decade or more, the formation histories and clustering
properties of dark matter halos, within a Lambda cold dark matter
($\Lambda$CDM) cosmology, have been studied in detail using both
numerical simulations and analytical analyses. These theoretical
studies have repeatedly shown that more massive halos are more
strongly clustered \citep[e.g.,][]{gang88, cole89, mo96, sheth99,
  wetzel07}, a result that is supported by the observed clustering of
nearby galaxy groups \citep{padilla04, yang05, berlind06,
  wang08}. Furthermore, recent dark matter simulations indicate that
the clustering of halos of a given mass depends on their assembly
history, where halos that assembled their mass earlier in the history
of the Universe are more strongly clustered than halos of the same
mass that formed later \citep{sheth04, gao05}; this effect is commonly
referred to as assembly bias \citep{croton07}.

The concept of assembly bias could be applicable to the formation of
galaxies in addition to dark matter halos, such that galaxies that
assembled their stellar mass earlier would be more strongly clustered
today than galaxies of like mass and younger stellar populations. Such
a bias in the galaxy population has typically been explored in two
distinct manners: [1] by studying galaxy samples spanning a range of
redshifts and directly observing the evolution in the population as a
function of mass and clustering properties, and [2] by applying a more
archaeological or paleontological approach, where the stellar
populations of nearby galaxies are studied in detail, with the goal of
disentangling the typical evolutionary history in different
environments.

Various observational studies have investigated the formation of
early--type or red--sequence galaxies from both perspectives. For
instance, analyses of samples drawn from the DEEP2 Galaxy Redshift
Survey \citep{davis03, newman09} by \citet{bundy06} and by
\citet{cooper06, cooper07} showed that the population of galaxies on
the red sequence at $z < 1$ was preferentially built--up in overdense
environments, so that at a given mass red galaxies in dense environs
have typically ceased their star formation (i.e., assembled their
mass) earlier than those in less dense regions. This picture is
supported by analyses of the fundamental plane \citep[FP,][]{dd87,
  dressler87} at similar redshifts, where it has been found that
early--type galaxies in high--density regions have reached the FP more
quickly than those in lower--density environs \citep{vandokkum01,
  gebhardt03, treu05, moran05}.

In contrast to the relatively coherent picture derived from studies of
early--type or red--sequence galaxies at intermediate redshift,
detailed studies of the stellar populations in nearby ellipticals have
yielded mixed results. A recent comparison of the stellar ages of
early--type systems in the Coma cluster versus a corresponding field
sample by \citet{trager08} found no difference in assembly history
between the two environment regimes. On the other hand, a similar analysis
of elliptical and lenticular galaxies in the Virgo and Coma clusters
by \citet{thomas05} found a distinct relationship between mean stellar
age and the environment in which a galaxy resides, such that cluster
early--types are older than their field counterparts.

In this paper, we take a different approach towards studying the topic
of assembly bias, with hopes of reconciling the results from studies
at intermediate--redshift with those focusing on local samples. Many
previous efforts at low redshift relied on comparing cluster and field
populations and/or focused on high--quality observations of a
relatively small number of galaxies and/or partially controlled for
morphology while ignoring inter--sample variations in properties such
as rest--frame color. In contrast, we utilize somewhat lower--quality
data for a much larger sample of galaxies drawn from the Sloan Digital
Sky Survey \citep[SDSS,][]{york00} to study the assembly histories of
thousands of red galaxies in a broad and continuous range of
environments, controlling for correlations between environment and
color, morphology, mass, etc. In the following Section (\S
\ref{sec_data}), we describe the galaxy sample under study, including
measurements of galaxy environments and stellar ages. Our results
regarding the relationship between stellar mass, age, and environment
are then presented in \S \ref{sec_results}, followed by further
analysis, discussion, and a summary in \S \ref{sec_anal}, \S
\ref{sec_disc}, and \S \ref{sec_summary}, respectively. Throughout, we
employ a $\Lambda$CMD cosmology with $w = -1$, $\Omega_m = 0.3$,
$\Omega_{\Lambda} = 0.7$, and a Hubble parameter of $H_0 = 100\ h$ km
s$^{-1}$ Mpc$^{-1}$, unless otherwise noted.

\section{Data}
\label{sec_data}

With spectra and multi--band photometry for more than $500,000$
galaxies, the SDSS Data Release 4 \citep[DR4,][]{dr4ref} enables the
local density of galaxies at $z \lesssim 0.2$ to be measured over
nearly one quarter of the sky. We select a parent sample of $372,688$
galaxies from the SDSS DR4, as drawn from the New York University
Value--Added Galaxy Catalog \citep[NYU--VAGC,][]{blanton05b}. This
sample is drawn from the NYU--VAGC Large--Scale Structure (LSS)
catalog, which provides a relatively uniform spatial selection
function by combining the SDSS imaging, target, and tiling masks while
tracking the flux limit and completeness as a function of angular
position. Our parent sample is limited to the redshift range $0.01 < z
< 0.3$ and to SDSS fiber plates for which the redshift success rate
for targets in the main spectroscopic survey is 80 percent or greater.

Rest--frame colors, absolute magnitudes, and stellar masses are
computed using the KCORRECT $K$--correction code (version v4\_1\_4) of
\citet[][see also \citealt{blanton03a}]{blanton07}. The rest--frame
quantities for the SDSS samples are derived from the SDSS apparent
$ugriz$ model magnitudes, where all magnitudes in this paper are
calibrated to the AB system \citep{oke83}. Stellar masses are computed
using template spectral energy distributions (SEDs) based on those of
\cite{bc03}. The best--fitting SED, given the observed $ugriz$
photometry and spectroscopic redshift, is used to directly compute the
stellar mass--to--light ratio $({\rm M}_{*}/L)$, assuming a
\citet{chabrier03} initial mass function.

\begin{figure}[h]
\centering
\plotone{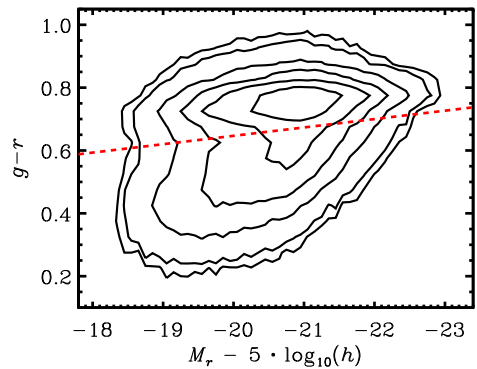}
\caption{We plot the rest--frame $g-r$ versus $M_r$ color--magnitude
  distribution for SDSS galaxies in the spectroscopic sample within
  the redshift range $0.05 < z < 0.15$. Due to the large number of
  galaxies in the sample, we plot contours (rather than individual
  points) corresponding to 25, 50, 200, 500, 1000, and 2000 galaxies
  per bin of $\Delta (g-r) = 0.05$ and $\Delta M_r = 0.1$. The dashed
  red horizontal line shows the division between the red sequence and
  the blue cloud as given in Equation \ref{eqn_color_cut}.}
\label{cmd_fig}
\end{figure}

As shown by many previous studies at low and intermediate redshift
\citep[e.g.,][]{strateva01, baldry04, bell04, willmer06}, and as
illustrated in Figure \ref{cmd_fig}, the distribution of galaxies in
color--magnitude space is bimodal, with a relatively tight red
sequence and a more diffuse blue cloud. To isolate the red--sequence
population, we use the following magnitude--dependent cut:
\begin{equation}
g - r = -0.02667 \cdot M_{r} + 0.11333.
\label{eqn_color_cut}
\end{equation}
This division in $g-r$ color is shown in Figure \ref{cmd_fig} as the
dashed red line.

To quantify stellar metallicities ($Z)$ and ages $(t)$ along the red
sequence, we employ the measurements of \citet{gallazzi05}, which are
based on model fits to spectral absorption features in the SDSS DR4
spectra. The model spectra utilized by \citet{gallazzi05} are derived
from the population synthesis models of \citet{bc03} and span a broad
range of star--formation histories. The models are simultaneously fit
to a minimum set of metal-- and age--sensitive spectral indices
(D$4000$, H$\beta$, H$\delta_A$ + H$\gamma_A$, [Mg$_2$Fe], and [MgFe],
\citealt{faber85,worthey94}), yielding measurements of age and
metallicity with typical uncertainties on the order of $\sigma_t
\lesssim 0.15$ dex and $\sigma_Z \lesssim 0.3$ dex, respectively.

While both the age and metallicity measurements are sensitive to the
signal--to--noise ratio of the galaxy spectrum \citep[see Table 1
of][]{gallazzi05}, we select our sample down to a relatively--faint
magnitude limit of $r = 17.5$. We investigate the sensitivity of our
results to this apparent $r$--band magnitude limit in \S
\ref{sec_anal}. Among the $114,916$ galaxies with $r < 17.5$ on the
red sequence at $0.05 < z < 0.15$, more than $93\%$ have robust age
and metallicity measures. The $\sim 7\%$ excluded from our sample
populate specific (and narrow) redshift windows, where critical age--
or metallicity--constraining spectral features (e.g., H$\beta$ or
H$\delta$) fall on sets of bright sky emission lines. Thus, the
galaxies with (or without) age and metallicity information are not
biased with regard to assembly history (i.e., not biased in terms of
stellar mass, morphology, etc.).

In addition to the stellar mass estimates discussed above, we also
employ the stellar mass measurements of \citet{gallazzi05}, which
differ from those computed using KCORRECT in that they do not rely on
fitting SEDs to the SDSS photometry; instead, the stellar masses from
\citet{gallazzi05} are a product of the same model fits to the SDSS
spectra that are used to constrain stellar age and metallicity, with
the SDSS model apparent magnitudes used only to normalize the
resulting stellar mass estimates. Thus, these additional stellar mass
values are complementary to those derived from KCORRECT while also
being congruous with the age and metallicity values employed in our
analysis. Note that the stellar masses of \citet{gallazzi05} utilize a
\citet{chabrier03} initial mass function, the same as that used by
KCORRECT.

In the SDSS, we estimate the local galaxy overdensity, or
``environment'' (as we will refer to it in this paper), using
measurements of the projected fifth--nearest--neighbor surface density
$(\Sigma_5)$ about each galaxy, where the surface density depends on
the projected distance to the fifth--nearest neighbor, $D_{p,5}$, as
$\Sigma_5 = 5 / (\pi D_{p,5}^2)$. Over quasi--linear regimes, the mass
density and galaxy density should simply differ by a factor of the
galaxy bias \citep{kaiser87}. In computing $\Sigma_5$, only objects
within a velocity window of $\pm1500$ km s$^{-1}$ are counted, to
exclude foreground and background galaxies along the line of sight.

To correct for the redshift dependence of the sampling rate of the
SDSS, each surface density is divided by the median $\Sigma_5$ for
galaxies within a window of $\Delta z = 0.02$ centered on the redshift
of each galaxy; this converts the $\Sigma_5$ values into measures of
overdensity relative to the median density (given by the notation $1 +
\delta_5$ herein) and effectively accounts for the redshift variations
in the selection rate \citep{cooper05}. We restrict our analyses to
the redshift range $0.05 < z < 0.15$, avoiding the low-- and
high--redshift tails of the SDSS galaxy redshift distribution where
the variations in the survey selection rate are greatest. Finally, to
minimize the effects of edges and holes in the SDSS survey geometry,
we exclude all galaxies within $1\ h^{-1}$ Mpc (comoving) of a survey
boundary, reducing our sample size to $102,601$ red galaxies within
the redshift range $0.05 < z < 0.15$. Note that all galaxies included
in the NYU--VAGC LSS sample (independent of color and down to the $r <
17.77$ spectroscopic limit) are used to trace the local galaxy
environment.

\section{Results}
\label{sec_results}

A wide array of galaxy properties at low and intermediate redshift
have been shown to correlate with environment. For example, at $z <
1$, there is a strong color--density relation such that red galaxies
tend to reside in higher density environs than their blue counterparts
\citep[e.g.,][]{hogg03, balogh04, cooper06}. Furthermore, within the
red galaxy population, there is also a significant luminosity
dependence to the clustering of galaxies, with the most luminous
members of the red sequence residing in the most massive dark matter
halos \citep[e.g.,][]{norberg02, hogg04, zehavi05, cooper06,
  coil06}. This relationship between environment, color, and
luminosity is such that at fixed color and luminosity no significant
trend is observed between local galaxy density and a variety of other
galaxy properties such as surface brightness, S\'{e}rsic index, or
stellar mass \citep{blanton05a, cooper08b}.

\begin{figure}[h]
\centering
\plotone{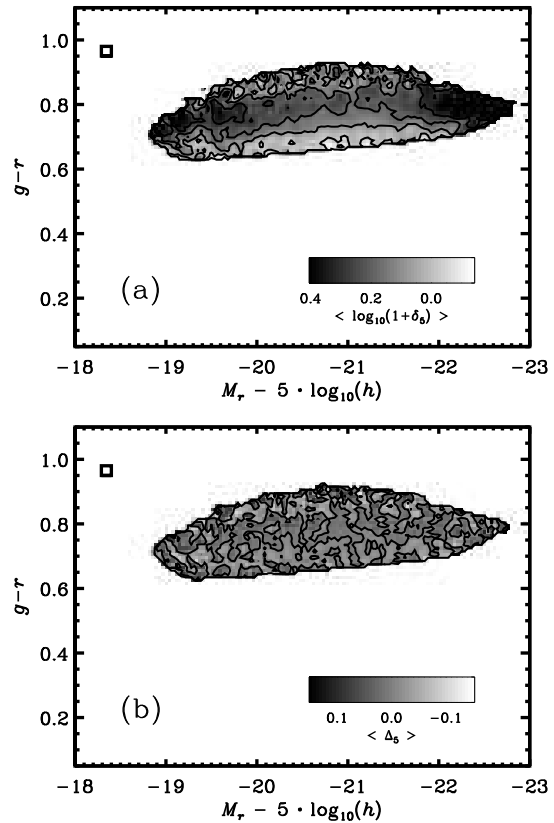}
\caption{(a) For the population of galaxies on the red sequence, we
  show the mean galaxy overdensity, $<\log_{10}(1+\delta_5)>$, as a
  function of rest--frame color, $g-r$, and absolute magnitude, $M_r$,
  computed in a sliding box of height $\Delta(g-r) = 0.03$ and width
  $\Delta M_r = 0.11$. The size and shape of the box are illustrated
  in the upper left--hand corner of the plot. (b) We plot the mean
  residual environment, $< \! \Delta_5 \!>$, as a function of color
  and magnitude, computed in the same sliding box.}
\label{cmdfit_fig}
\end{figure}

However, a recent analysis of the relationship between environment and
the luminosities, rest--frame colors, stellar masses, and gas--phase
metallicities of nearby star--forming galaxies selected from the SDSS
by \citet{cooper08b} shows that after removing the mean dependence of
environment on color and luminosity there remains a
highly--significant correlation between environment and oxygen
abundance. That is, they find that within the star--forming population
(i.e., on the blue cloud) there is a relationship between local galaxy
density and gas--phase metallicity separate from that observed between
galaxy density and color and luminosity (or stellar
mass). Equivalently, \citet{cooper08b} show that at fixed stellar mass
and star--formation rate, star--forming galaxies that reside in more
overdense environments tend to have more metal--rich gas in their
H{\small II} regions.

In this paper, we apply the same techniques to study the relationship
between mean stellar age and environment at fixed stellar mass along
the red sequence. Following \citet{cooper08b}, we first measure and
remove (subtract) the mean dependence of environment on color and
luminosity along the red sequence. Figure \ref{cmdfit_fig}a shows the
mean overdensity as a function of rest--frame $g-r$ color and
$r$--band absolute magnitude, $< \log_{10}(1+\delta_5)[g-r,M_r] >$,
for our sample of red galaxies. In addition to a relatively weak
color--density trend (when restricting to red galaxies), we find a
strong luminosity--density relation, with the mean overdensity
increasing at the under-- and over--luminous ends of the red sequence,
as observed by \citet{hogg03} and \citet{blanton05a}. To remove this
relationship of environment to color and luminosity, we subtract the
mean overdensity at the color and luminosity of each galaxy from the
measured overdensity:
\begin{equation}
\Delta_5 = \log_{10}(1+\delta_5) - < \log_{10}(1+\delta_5)[g-r,M_r] > ,
\label{eqn_delta5}
\end{equation}
where the distribution of mean environment with color and absolute
magnitude, $< \log_{10}(1+\delta_5)[g-r,M_r] >$, is median smoothed on
$\Delta (g-r) = 0.09$ and $\Delta M_r = 0.33$ scales prior to
subtraction. Using rest--frame $u-r$ color in lieu of $g-r$ creates a
greater dynamic range in color along the red sequence, but
precipitates no change in our results regarding the relationships
between environment and age, metallicity, and stellar mass.

The ``residual'' environment, $\Delta_5$, quantifies the overdensity
about a galaxy relative to galaxies of similar color and luminosity,
where values of $\Delta_5$ greater than zero correspond to galaxies in
environments more overdense than that of the typical galaxy with like
color and luminosity (i.e., with like star--formation history). Figure
\ref{cmdfit_fig}b shows the dependence of mean $\Delta_5$ on color and
luminosity; no significant color or luminosity dependence is evident,
with deviations from $< \! \Delta_5 \! > \: = 0$ being small. For more
details and discussion of this technique, refer to
\citet{cooper08b}. Finally, note that the ``residual'' overdensity,
$\Delta_5$, is only a small perturbation to the ``absolute''
overdensity, $\log_{10}(1+\delta_5)$, such that the two environment
statistics are still strongly correlated with each other for
individual objects.

\begin{figure*}[tb]
\centering
\plotone{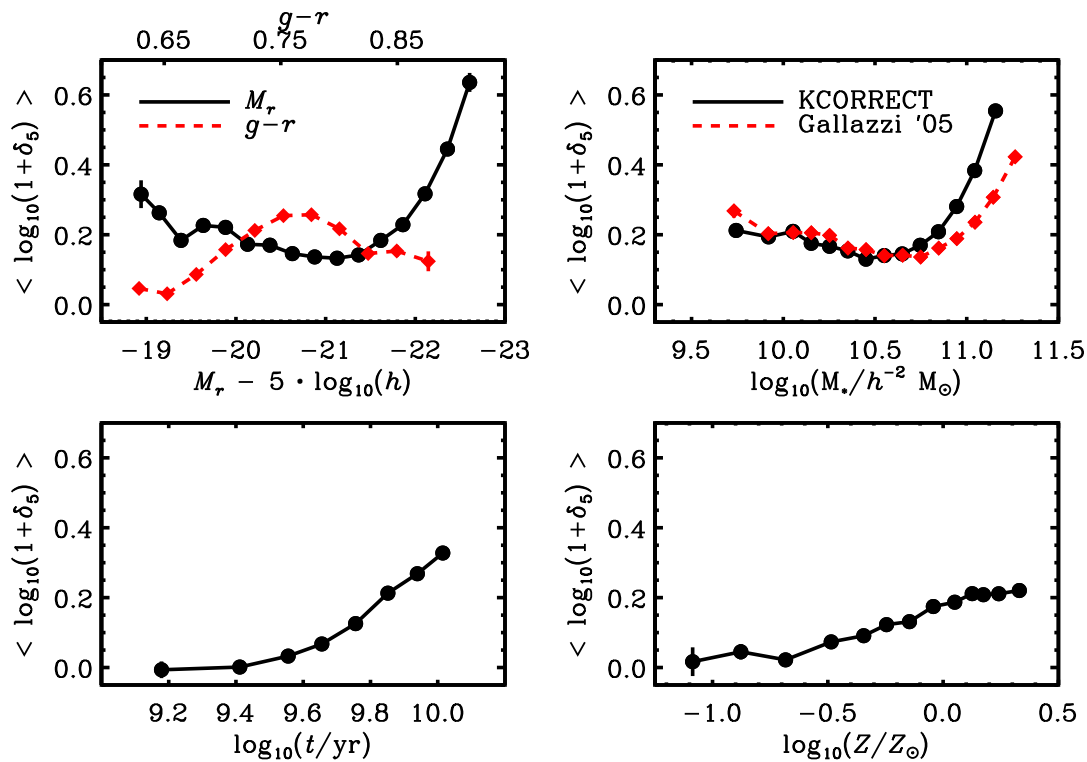}
\caption{The dependence of mean overdensity,
  $<\log_{10}(1+\delta_5)>$, (i.e., ``absolute'' environment) on the
  absolute magnitude, rest--frame color, stellar mass, stellar age,
  and stellar metallicity of red--sequence galaxies. The points and
  error bars give the mean environments and $1\sigma$ uncertainties in
  the means computed in distinct bins. In the top right panel, we show
  the results based on two independent stellar mass estimates, those
  determined solely from fits to the broad--band photometry using
  KCORRECT and those based on model fits to the SDSS spectra from
  \citet{gallazzi05}. As seen in previous studies, we find strong
  luminosity--density and mass--density relations along the red
  sequence. We also observe relatively strong correlations between
  average environment and other properties of the stellar population,
  such as age and metallicity.}
\vspace*{0.1in}
\label{absolut_fig}
\end{figure*}

In Figure \ref{absolut_fig}, we show the average dependence of
absolute environment, $\log_{10}(1+\delta_5)$, on luminosity,
rest--frame color, stellar mass, stellar age, and stellar metallicity
for galaxies on the red sequence. As seen in many previous studies, we
find a strong relationship between luminosity and local galaxy
density, driven by luminous galaxies at the centers of groups and
clusters and fainter satellite galaxies sitting in the outskirts of
overdense regions \citep{berlind05}. The color--density relation is
relatively weak over the narrow range of rest--frame colors spanned by
the red sequence. The small measured variations in the average
overdensity with color are likely associated with the projection of
the luminosity--density relation that results from the intrinsic tilt
of the red sequence towards redder colors at brighter luminosities.
In addition, the observed fluctuations in the color--density relation
may be caused by interloping disk galaxies, which appear on the red
sequence due to dust and inclination effects \citep[e.g.,][]{weiner05,
  driver06, lotz08, maller09}; away from the ridge of the red
sequence, these interlopers have a stronger impact on the average
environment, due to their greater relative numbers. Finally, the
mass--environment relationship within our sample closely follows that
between luminosity and environment, as expected given the strong
correlation between the combination of $g-r$ and $M_r$ with stellar
mass \citep[e.g.,][]{bell01, kauffmann03, cooper08a}.

In addition to strong luminosity-- and mass--environment trends along
the red sequence, we find a significant relationship between mean
absolute environment and luminosity--weighted mean stellar age, where
galaxies with older stellar populations favor overdense regions (cf.\
Figure \ref{absolut_fig}). With a typical uncertainty of $\sigma_t
\lesssim 0.15$ dex for each individual age measurement \citep[see
Table 1 of ][]{gallazzi05}, the residual age--density relation spans
multiple independent bins, while also covering nearly $0.4$ dex in $<
\! \Delta_5 \! >$.\footnote{The median uncertainties in the age and
  metallicity measurements for the $r < 17.5$ galaxy sample are
  $\sigma_{t} = 0.11$ dex and $\sigma_{Z} = 0.13$ dex, respectively.}

We also find a significant trend between stellar metallicity, $Z$, and
absolute environment along the red sequence, such that more
metal--rich galaxies favor higher--density regions. However, it should
be noted that there is a non--negligible correlation between the
binned data points in Fig.\ \ref{absolut_fig}, given the relatively
large uncertainty in the typical metallicity measurement, $\sigma_{Z}
\sim 0.1$--$0.5$ dex. Furthermore, the vast majority $(\gtrsim \!
75\%)$ of the red galaxy population has measured metallicities of
$\log_{10}(Z/Z_{\sun}) \gtrsim -0.1$, so the apparent strength of the
metallicity--density relation is driven in part by a relatively small
number of galaxies in the sample and by those systems with the
greatest uncertainty in $Z$ --- the $\sigma_Z$ values are typically
larger at low metallicities, with a significant tail to $\sigma_z \sim
0.5$ dex. Altogether, the strength of the metallicity--density
relation is considerably weaker than that of the luminosity--, stellar
mass--, or age--density relations, though still highly statistically
significant.

\begin{figure*}[tb]
\centering
\plotone{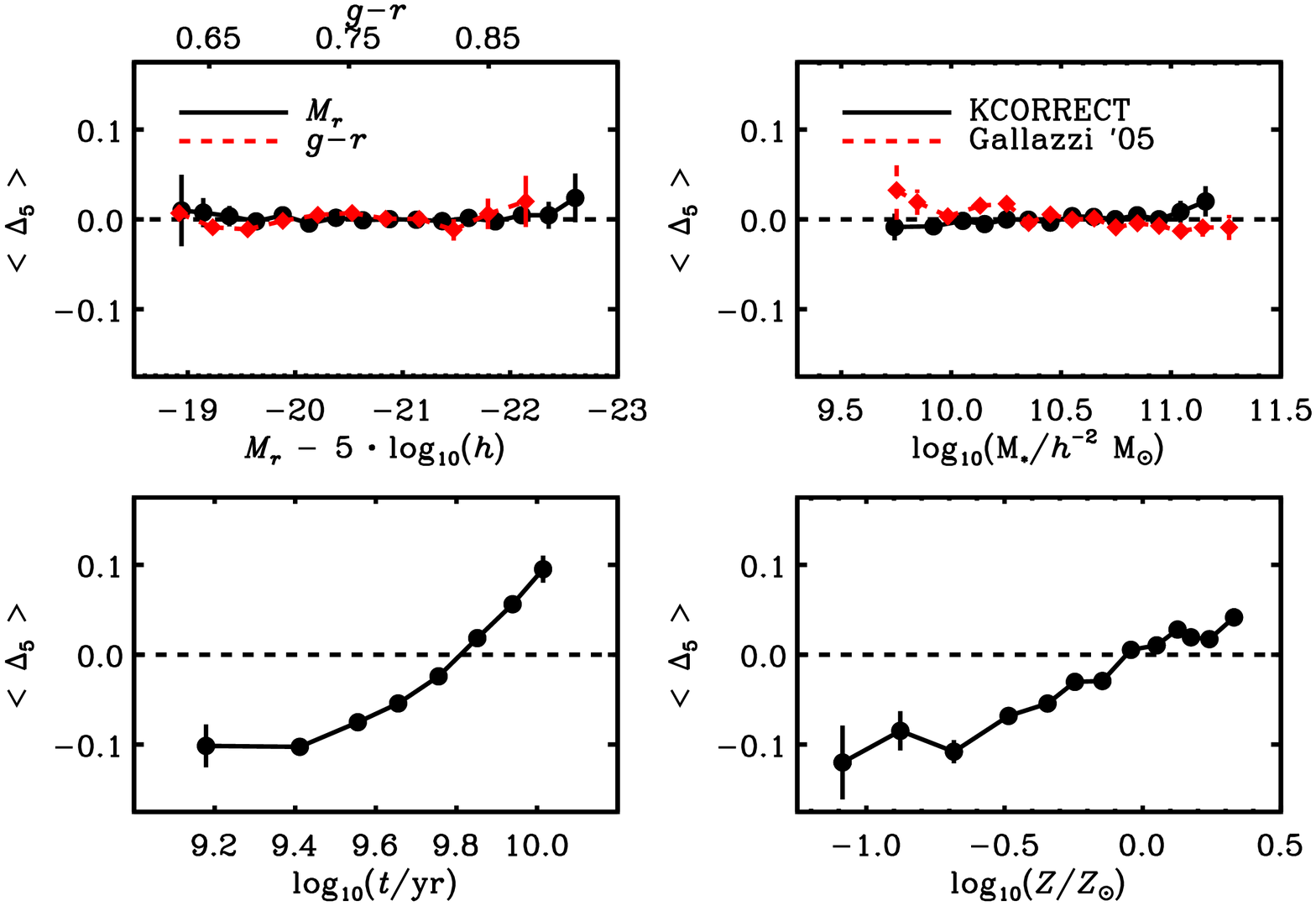}
\caption{The dependence of mean residual environment, $< \! \Delta_5
  \! >$, on absolute magnitude, rest--frame color, stellar mass,
  stellar age, and stellar metallicity. The points and error bars
  correspond to the mean environments and $1\sigma$ uncertainties in
  the means computed in distinct bins of each galaxy property. In the
  top right panel, we show the results based on two independent
  stellar mass estimates, those determined solely from fits to the
  broad--band photometry using KCORRECT and those based on model fits
  to the SDSS spectra from \citet{gallazzi05}. While we find no
  significant residual trend between environment and color,
  luminosity, or stellar mass, we observe a strong dependence of
  residual environment on stellar age and metallicity.}
\vspace*{0.1in}
\label{resid_fig}

\end{figure*}

By studying the dependence of residual environment, $\Delta_5$, on
each galaxy characteristic (e.g., $M_*$, $t$, and $Z$), we can
determine whether any particular property exhibits an excess
correspondence with local galaxy density beyond that explained by the
color--luminosity--environment relation. In Figure \ref{resid_fig}, we
show the average dependence of mean residual environment on each
galaxy property shown in Figure \ref{absolut_fig} (color, luminosity,
stellar mass, age, and metallicity). By construction, there is no
significant correlation between average $\Delta_5$ and luminosity or
rest--frame color (cf.\ Table \ref{tab1}). Similarly, we find no
dependence of mean residual environment on stellar mass within the
red--sequence population for both sets of stellar mass measurements
described in \S \ref{sec_data}. The previously--noted tight
relationship between stellar mass and rest--frame color and luminosity
mass makes this result more or less expected.

While we find no significant correlation between residual environment
and color, luminosity, or stellar mass among red galaxies in the SDSS,
we do see a strong relationship between $\Delta_5$ and stellar age,
such that galaxies with greater stellar ages tend to reside in regions
of higher density relative to other galaxies of like color and
luminosity (i.e., of like stellar mass). In other words, we find that
at a given stellar mass red galaxies with older stellar populations
favor overdense environments relative to their younger counterparts.

Likewise, we find a significant correlation between stellar
metallicity, $Z$, and residual environment; this residual
metallicity--density trend is such that more metal--rich galaxies
favor more overdense regions than galaxies of like color and
luminosity. The residual age--density and metallicity--density
relations are quite strong, spanning roughly $0.2$ dex and $\gtrsim
0.1$ dex in $\Delta_5$, respectively, or more than half the range in
overdensity spanned by the absolute age--density and
metallicity--density relations. However, as stated previously, the
strength of the metallicity--density trend is driven by a relatively
small number of galaxies at low metallicities $(\log_{10}(Z/Z_{\sun})
\lesssim -0.1)$; in the more metal--rich regime, where the bulk of our
sample resides, the residual metallicity--density relation is
relatively weak.

\begin{deluxetable}{l c c c}
  \tablewidth{0pt} \tablecolumns{4}
  \tablecaption{\label{tab1} Statistical Strengths of
    Relations with $\Delta_5$} 
\startdata \hline
Galaxy Property & $\rho$ & slope & $\sigma_{\rm slope}$ \\
  \hline \hline
$M_r$ & -0.006 & 0.001 & 0.002 \\
$g-r$ & -0.001 & 0.046 & 0.109 \\
$\log_{10}({\rm M}_{*}/{\rm M}_{\sun})$ & 0.009 & 0.011 & 0.005 \\
$\log_{10}(t/{\rm yr})$ & 0.069 & 0.333 & 0.042 \\
$\log_{10}(Z/Z_{\sun})$ & 0.039 & 0.135 & 0.017 \\
\enddata
\tablecomments{We quantify the correlations between various galaxy
  properties and residual density $(\Delta_5)$ for the sample of
  red--sequence galaxies with $r < 17.5$. Here, we list the Spearman
  ranked correlation coefficient $(\rho)$ as well as the measured
  slope ($d<\! \Delta_5 \! >/dx$) and its uncertainty derived from
  linear fits to the binned data points in Figure \ref{resid_fig}. The
  statistical strengths given in the table for the stellar
  mass--$\Delta_5$ relation correspond to the stellar mass values
  produced by the KCORRECT code.}
\end{deluxetable}

\section{Potential Systematic Effects}
\label{sec_anal}

In \S \ref{sec_results}, we presented the relationships between (both
absolute and residual) environment and an assortment of galaxy
properties, including $r$--band luminosity, rest--frame $g-r$ color,
stellar mass, luminosity--weighted mean stellar age, and stellar
metallicity for a sample of red galaxies with $r < 17.5$. While we
find no correlation between mean residual environment and stellar
mass, we do observe significant relationships between $< \! \Delta_5
\! >$ and stellar age and metallicity. In the remainder of this
section, we discuss possible systematic effects that could bias these
results.

\subsection{Sample Selection and Aperture Effects}
\label{sec_disc1}

In contrast to other studies of the relationship between environment
and stellar age or metallicity for early--type galaxies in the local
Universe \citep[e.g.,][]{trager08}, our work relies upon estimates of
age and metallicity derived from relatively low--quality (i.e., low
signal--to--noise ratio) spectra, which translates into measurements
with greater uncertainties. It is only by constructing large samples
that we are able to overcome these greater uncertainties and conduct
detailed analyses of the mean relations with environment.

Given the larger uncertainty in each individual age and metallicity
measurement, it is important that we clearly understand any potential
bias in our sample selection. The errors in the estimates of both $Z$
and $t$ by \citet{gallazzi05} depend on the signal--to--noise ratio of
the SDSS spectra, with metallicity measurements showing greater
sensitivity to the quality of the data. \citet{gallazzi05} find that
the uncertainty in the estimated age shows less variation with the
mean signal--to--noise ratio per pixel of the SDSS spectrum than the
uncertainty in the metallicity constraint (cf.\ Section 2.4.1 and
Figure 6 of \citealt{gallazzi05}). This is attributed to the fact that
stellar age is largely constrained by the $4000$\AA\ break and the
Balmer absorption lines, while metallicity estimation relies heavily
on Mg and/or Fe features, which are more difficult to measure at a
given signal--to--noise ratio, especially in low--mass/metal--poor
systems where these features are commonly weak.

\begin{figure}[h]
\centering
\plotone{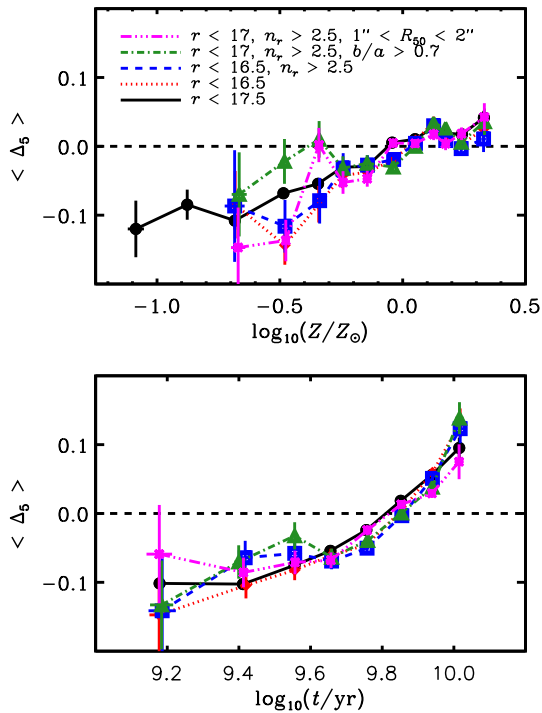}
\caption{The dependence of mean residual environment, $< \! \Delta_5
  \!>$ on stellar metallicity (top) and stellar age (bottom), for a
  variety of sample selections. The black circles connected by the
  solid black line correspond to the main $r < 17.5$ red galaxy
  sample. The relations for three more--selective samples are
  overplotted: a $r < 16.5$ magnitude--limited sample corresponds to
  the red diamonds connected by the dotted red line, a $r < 16.5, n_r
  > 2.5$ sample corresponds to the blue squares connected by the
  dashed blue line, and a $r < 17, n_r > 2.5, b/a > 0.7$ sample
  corresponds to the green triangles connected by the dashed--dotted
  green line. Even when including strong morphological, color, and
  brightness limits, the residual age--density relation persists.}
\label{select_fig}
\end{figure}

To test the dependence of our results on the uncertainty in the age
and metallicity estimates (i.e., on the quality of the spectra), we
vary the magnitude limit utilized in the sample selection, defining
brighter subsamples for which the typical spectrum has a higher
signal--to--noise ratio (cf.\ \S \ref{sec_data}). In addition to
utilizing a $r < 17.5$ limit, we study samples restricted to have $r <
17$ or $r < 16.5$. Across all of these samples, our general results
regarding the relationship between stellar age and environment remain
unchanged; we still find a highly significant residual age--density
relation in each of the magnitude--limited samples (cf.\ Figure
\ref{select_fig}), while no correlation is found between average
$\Delta_5$ and color, luminosity, or stellar mass. Furthermore, the
strength of the residual age--density relation remains unchanged, even
when limiting the sample to only those galaxies with higher--quality
spectra (cf.\ Figure \ref{select_fig} and Table \ref{tab2}).

While our results regarding the age--environment relation are very
robust to variations in the apparent magnitude limit of the sample,
the relationship between stellar metallicity and environment becomes
increasingly less significant as the sample is restricted to brighter
(and therefore preferentially more luminous and more massive)
systems. In particular, the range in stellar metallicity probed by the
$r < 17$ and $r < 16.5$ samples is reduced to $\log_{10}(Z/Z_{\sun})
\gtrsim -0.75$ and the uncertainties in the mean residual environment,
$< \! \Delta_5 \! >$, increase such that the resulting correlation
between average residual environment and stellar metallicity is only
statistically significant within the relatively metal--rich regime
(i.e., $\log_{10}(Z/Z_{\sun}) \gtrsim -0.35$).

In addition to concerns regarding the impact of low--quality data, our
results could be biased by the inclusion of dusty, edge--on
star--forming galaxies in our analysis, which would skew the galaxy
sample towards lower stellar ages in low--density environments. Dusty
star--forming systems may also be a concern in the higher--density
regime; recent results from the Space Telescope A901/902 Galaxy
Evolution Survey \citep[STAGES,][]{gray09, gallazzi09, wolf09} and
the Local Cluster Substructure Survey \citep[LoCuSS,][]{haines09} have
suggested that obscured star formation may be prevalent in
higher--density environments such as the outskirts of clusters. As
highlighted earlier in this paper, analyses of galaxy samples at
intermediate and low redshift have shown that some galaxies on the red
sequence are not truly ``red and dead'', early--type systems, but are
dusty (i.e., highly reddened) star--forming disk galaxies
\citep[e.g.,][]{bell04b, weiner05}. This interloper population
comprises an increasingly greater fraction of the red sequence at
lower luminosities \citep[e.g.,][]{lotz08, maller09}, so restricting
our analysis to brighter magnitude limits, as discussed above, would
be expected to reduce the impact of dusty disk galaxies.

However, restricting our sample according to morphology provides a
more robust means for minimizing the contribution of dusty disk
galaxies to the measured age--density relation. Utilizing the
$r$--band S\'{e}rsic indices, $n_r$, of \citet{blanton03b}, we select
from our red galaxy samples those sources with $n_r > 2.5$. For this
color-- and morphology--selected sample, the residual relationship
between environment and stellar age remains strong.  Even when
restricting our analysis to $r < 16.5$ and $n_r > 3$, we still find no
significant change in the strength of the residual age--density
relation, though the uncertainties are greater due to the smaller
sample size.

To improve morphological discrimination further, we also employ
measurements of the $r$--band isophotal major and minor axes from the
SDSS imaging to remove highly--inclined disk galaxies. Restricting to
sources with an axis ratio $b/a > 0.7$, with or without a joint
morphological selection $(n_r > 2.5)$, we continue to find a strong
correlation between mean stellar age and residual environment (cf.\
Figure \ref{select_fig} and Table \ref{tab2}). Furthermore, we also
observe no significant dependence of mean $\Delta_5$ on $n_r$ or
$b/a$, indicating that the observed correlation between stellar age
and residual environment is not reflecting an unaccounted--for
morphological dependence.

While the age--density relation is quite robust to additional
selection criteria, the relationship between residual environment and
stellar metallicity becomes increasingly weaker and less significant
as the galaxy sample is restricted according to color, apparent
magnitude, and morphology. As shown in Figure \ref{select_fig},
when limiting our sample to those galaxies with $r < 17$, $n_r > 2.5$,
and $b/a > 0.7$, we find weaker (though still statistically
significant) evidence of a residual metallicity--density relation,
with an average residual environment of $< \! \Delta_{5} \! > \: \sim
0$ across the entire range of metallicities probed.

A second way to test the robustness of our results to contamination on
the red sequence is to make our magnitude--dependent color cut more
restrictive, which will reduce the number of galaxies in the ``green
valley'' or the red extremes of the blue cloud that are included in
our sample. In addition to the brighter $r$--band magnitude limit of
$r < 16.5$ and the S\'{e}rsic selection cut of $n_r > 3$, we shift our
color selection, as given in Equation \ref{eqn_color_cut}, redward by
$g-r = 0.05$. For this restricted sample of bright, red,
centrally--concentrated galaxies, we continue to find a strong
residual age--density relation, spanning $\sim 0.2$ dex in $< \!
\Delta_{5} \! >$, while the metallicity--density relation is
diminished --- although a significant residual relationship persists
between metallicity and environment beyond that observed with color
and luminosity.

Another potential systematic effect to consider is the impact of
aperture bias on our analysis. At $z = 0.1$, an SDSS fiber (3 arcsec
in diameter) subtends a physical distance of only $\sim 5.5$ kpc
(assuming a Hubble constant of $H_0 = 70$ km s$^{-1}$
Mpc$^{-1}$). Thus, the \citet{gallazzi05} metallicities and ages are
estimates for the central portion of each galaxy. However, as
discussed in more detail by \citet{gallazzi05}, there is no apparent
dependence of mean stellar age on galaxy concentration or redshift, at
fixed luminosity within their catalog. The lack of any significant
aperture bias is additionally supported by detailed studies of
early--type galaxies in the local Universe \citep[e.g.,][]{mehlert03,
  wu05, sanchez07}, which find no significant age gradient out to
large physical radii. As a secondary check, we restrict our sample to
those sources with an $r$--band half--light radius between $1$ and $2$
arcseconds as well as $r < 17$ and $n_r < 2.5$. The residual
age--density and metallicity--density relations for this subsample
remain highly significant, as shown in Table \ref{tab2}. These results
are consistent with a lack of aperture bias.

\begin{deluxetable*}{l c  c c c  c c c}
  \tablewidth{0pt} \tablecolumns{7}
  \tablecaption{\label{tab2} Statistical Strengths of
    Relations with $\Delta_5$ for Different Galaxy Subsamples} 
\startdata \hline
  \multirow{2}{*}{Sample} & \multirow{2}{*}{$N_{\rm galaxy}$} & 
  \multicolumn{3}{c}{$\log_{10}(t/{\rm yr})$} & 
  \multicolumn{3}{c}{$\log_{10}(Z/Z_{\sun})$} \\
  & & $\rho$ & slope & $\sigma_{\rm slope}$ & $\rho$ & slope &
  $\sigma_{\rm slope}$ \\
  \hline \hline
  $r < 17.5$                     
  & 104,745 & 0.069 & 0.333 & 0.042 & 0.039 & 0.135 & 0.017 \\
  $r < 16.5$                     
  & 30,446  & 0.076 & 0.343 & 0.081 & 0.029 & 0.143 & 0.057 \\
  $r < 16.5, n_r > 2.5$                     
  & 27,024  & 0.066 & 0.325 & 0.120 & 0.022 & 0.122 & 0.052 \\
  $r < 17, n_r > 2.5, b/a > 0.7$             
  & 29,002  & 0.049 & 0.288 & 0.096 & 0.030 & 0.085 & 0.033 \\
  $r < 17, n_r > 2.5, 1" < R_{50} < 2"$ 
  & 28,020  & 0.043 & 0.298 & 0.052 & 0.029 & 0.128 & 0.065 \\
\enddata
\tablecomments{We quantify the correlations between stellar age and
  metallicity and residual density $(\Delta_5)$ for the set of galaxy
  samples analyzed in Figure \ref{select_fig}. The number of galaxies
  $(N_{\rm galaxies})$ in each sample is given in the second
  column. In addition, we list the Spearman ranked correlation
  coefficient $(\rho)$ as well as the measured slope (and its
  uncertainty) derived from linear fits to the corresponding binned
  data points in Figure \ref{select_fig}.}
\end{deluxetable*}

\subsection{Stellar Mass versus Velocity Dispersion}

While the topic of assembly bias, as applied towards galaxy assembly
versus halo assembly in this work, focuses on the relationship between
galaxy assembly time and clustering at fixed assembled mass (i.e.,
fixed stellar mass), it is also common and interesting to study
differences in galaxy clustering as a function of age at fixed halo
mass or fixed total (dark plus baryonic) mass. For instance, many
studies of nearby samples in clusters and the field
\citep[e.g.,][]{thomas05, clemens06, trager08} compare the ages of
stellar populations in early--type galaxies at fixed stellar velocity
dispersion (a common proxy for halo mass).

Here, we employ the velocity dispersion measurements generated by the
Princeton SDSS spectral reduction pipeline (Schlegel \& Burles in
prep). The velocity dispersion, $\sigma$, for each galaxy is estimated
via the ``direct--fitting'' method \citep{burbidge61, rix92}, in which
the best--fit $\sigma$ is determined by comparing (in pixel space) a
Gaussian--broadened stellar template to the observed galaxy spectrum
and fitting for the reduced--$\chi^2$ minimum.

\begin{figure}[h]
\centering
\plotone{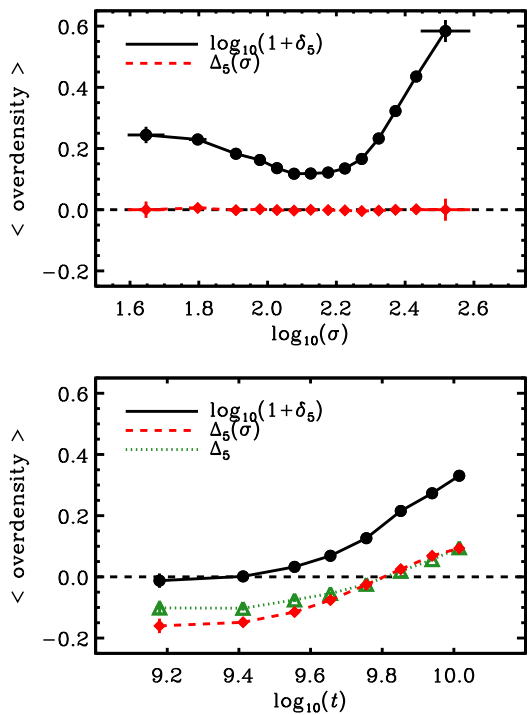}
\caption{The dependence of mean absolute and residual environment on
  velocity dispersion (top) and stellar age (bottom), for the sample
  of red--sequence galaxies with $r < 17.5$. The black circles
  connected by the solid black line correspond to the dependence of
  mean absolute environment, $\log_{10}(1+\delta_5)$, on velocity
  dispersion or stellar age, while the red diamonds connected by the
  dashed red line give the respective relationship with residual
  environment, $\Delta_5(\sigma)$. The green triangles connect by the
  dotted line in the bottom panel show the average relationship
  between residual environment, $\Delta_5$, and age, as shown in
  Figure \ref{resid_fig}. Even when substituting stellar velocity
  dispersion for stellar mass, the strong residual age--density
  relation persists.}
\label{vdisp_fig}
\end{figure}

As shown in Figure \ref{vdisp_fig}, velocity dispersion has a strong
relationship with environment, similar to the observed luminosity--
and stellar mass--environment relations (cf.\ Fig.\ \ref{absolut_fig})
--- the more massive systems are commonly found in increasingly
higher--density regions, while the average density also increases at
low velocity dispersion largely due to lower--mass satellite galaxies,
which reside in more massive (i.e., more strongly--clustered) halos.
The velocity dispersion measurements utilized in Figure
\ref{vdisp_fig} are not aperture--corrected. However, applying
standard corrections \citep[e.g.,][]{davies87, jorgensen95} yields no
significant changes in the strengths of the observed relations with
environment.

To remove the relationship between environment and stellar velocity
dispersion, we subtract the mean overdensity at the velocity
dispersion of each galaxy from the measured overdensity:
\begin{equation}
  \Delta_5(\sigma) = \log_{10}(1+\delta_5) - < \log_{10}(1+\delta_5)[\log_{10}(\sigma)] > ,
\label{eqn_delta5_vdisp}
\end{equation}
yielding a new ``residual'' environment measure that traces the local
overdensity relative to galaxies of like velocity dispersion, but
possibly differing luminosity or rest--frame color.

In Figure \ref{vdisp_fig}, we show the average dependence of this new
residual environment, $\Delta_5(\sigma)$, on velocity dispersion and
stellar age for the $r < 17.5$ magnitude--limited sample of
red--sequence galaxies. By construction, $\Delta_5(\sigma)$ exhibits
no dependence on velocity dispersion (cf.\ Table \ref{tab3}), while a
strong residual age--density relation is found, following the same
dependence observed with $\Delta_5$; that is, we find that after
removing the dependence of environment on velocity dispersion, there
exists a highly--significant relationship between luminosity--weighted
mean stellar age and environment, such that galaxies with older
stellar populations favor regions of higher density relative to
galaxies of like velocity dispersion. Furthermore, the strength of the
age--$\Delta_5(\sigma)$ relation (as shown in Fig.\ \ref{vdisp_fig})
exceeds that of the age--$\Delta_5$ relation (as shown in Fig.\
\ref{resid_fig}) for the same sample of red galaxies (cf.\ Table
\ref{tab2} and Table \ref{tab3}). The relationship between stellar age
and environment at fixed velocity dispersion spans $\sim \! 0.25$ dex
in $\Delta_5(\sigma)$, while the relationship between age and
environment at fixed color and luminosity spans only $\sim \! 0.2$ dex
in $\Delta_5$. Thus, not only do we find a highly--significant
relationship between age and environment when controlling for the
dependence of galaxy density on velocity dispersion, but the residual
age--density relation is observed to be stronger at fixed velocity
dispersion than at fixed color and luminosity (i.e., at fixed stellar
mass).

\begin{deluxetable*}{l  c c c  c c c}
  \tablewidth{0pt} \tablecolumns{6}
  \tablecaption{\label{tab3} Statistical Strengths of
    Relations with $\Delta_5(\sigma)$ for Different Galaxy Subsamples} 
\startdata \hline
  \multirow{2}{*}{Sample} &
  \multicolumn{3}{c}{$\log_{10}(\sigma)$} &
  \multicolumn{3}{c}{$\log_{10}(t/{\rm yr})$} \\
  & $\rho$ & slope & $\sigma_{\rm slope}$ & $\rho$ & slope &
  $\sigma_{\rm slope}$ \\
  \hline \hline
  $r < 17.5$                     
  & -0.002 & -0.003 & 0.006 & 0.093 & 0.480 & 0.019 \\
  $r < 16.5$                     
  & 0.001  & -0.008 & 0.010 & 0.079 & 0.396 & 0.036 \\
  $r < 16.5, n_r > 2.5$          
  & -0.001 & -0.004 & 0.023 & 0.072 & 0.426 & 0.050 \\
  $r < 17, n_r > 2.5, b/a > 0.7$ 
  & 0.005  & 0.008  & 0.009 & 0.076 & 0.483 & 0.081 \\
  $r < 17, n_r > 2.5, 1" < R_{50} < 2"$ 
  & 0.000 & -0.017 & 0.019 & 0.073 & 0.467 & 0.041 \\
\enddata
\tablecomments{We quantify the correlations between velocity
  dispersion and stellar age and the residual density,
  $\Delta_5(\sigma)$, for the set of galaxy samples analyzed in Figure
  \ref{select_fig}. We list the Spearman ranked correlation
  coefficient $(\rho)$ as well as the measured slope (and its
  uncertainty) derived from linear fits to the binned data points
  tracing the $\Delta_5(\sigma)[\log_{10}(\sigma)]$ and
  $\Delta_5(\sigma)[\log_{10}(t)]$ relations, using the same discrete
  bins employed in Figure \ref{vdisp_fig}.}
\end{deluxetable*}

\subsection{Addressing the Age--Metallicity Degeneracy}
\label{sec_degen}

It has been understood for many years that the integrated colors and
spectra of stellar populations depend on both metallicity and
formation epoch (i.e., age) in a degenerate manner
\citep[e.g.,][]{worthey94, ferreras99}; increasing the metallicity of
a population at fixed age has a similar effect on integrated colors as
increasing the age at fixed metallicity. The use of more age--specific
and metallicity--specific spectral features, such as Balmer, Mg, and
Fe absorption features, has partially broken this degeneracy
\citep[e.g.,][]{trager98, faber99}. However, errors in estimating
stellar age (i.e., errors in measuring the strength of specific
features such as H$\beta$ absorption) remain strongly correlated with
errors in metallicity. For instance, if a galaxy's age is
overestimated, then its metallicity is correspondingly underestimated.

Even given the correlation between errors in metallicity and errors in
age, it is highly unlikely that our results are biased by such
effects. To start with, there is no physical mechanism that would
cause the \citet{gallazzi05} age estimates to be off in an
environment--dependent manner, thereby creating a bogus age--density
relation. Still, if we assume that the Balmer lines (H$\beta$ and
H$\delta$) used to constrain age by \citet{gallazzi05} are over-- or
under--estimated due to poor emission--line corrections, such that a
galaxy thought to be old is actually young, then that galaxy must also
be more metal--rich, given the relationship between age and
metallicity in stellar population models. Thus, even if errors in age
measurements conspire to cause the observed residual age--density
relation (a physically unmotivated scenario), any correction to the
age estimates would produce a strengthening of the residual
metallicity--density relation.

Overall, there is no reason to think that the \citet{gallazzi05}
measurements are off in an environment--dependent manner. However, as
always, age and metallicity estimates are only as good as the
theoretical models (i.e., the range of star--formation histories and
stellar population parameters probed by the models). Still, any
limitation or unprobed range of parameter space associated with the
models must be correlated with environment at fixed mass in order to
explain our results. Thus, our general conclusion that differences
exist in the typical stellar populations of nearby galaxies in
high--density and low-density environs cannot be explained in terms of
model deficiencies, as any limitations of the models (e.g., the
exclusion of blue horizontal--branch stars) must be correlated with
environment, which directly translates into a difference in stellar
populations with environment.

\section{Discussion}
\label{sec_disc}

In \S \ref{sec_results}, we presented the relationships between (both
absolute and residual) environment and an assortment of galaxy
properties, including $r$--band luminosity, rest--frame $g-r$ color,
stellar mass, luminosity--weighted mean stellar age, and stellar
metallicity. While we find no correlation between mean residual
environment and stellar mass, we do observe significant relationships
between $< \! \Delta_5 \! >$ and stellar age and metallicity. We
discuss the implications of these results in the remainder of this
section.

\subsection{Comparison to Previous Studies}

\subsubsection{Age--Density Relation}

As highlighted in \S \ref{sec_intro}, studies of red--sequence
galaxies in local clusters and within the field population have
yielded discrepant results on the correlation between age and
environment at fixed mass. On one hand, a variety of analyses studying
early--type galaxies (i.e., E/S0 morphological types) in nearby
clusters and the field have arrived at similar conclusions as our
work, where red galaxies in higher--density regions are typically
older at a given mass \citep[e.g.,][]{trager00b, thomas05, clemens06,
  bernardi06, smith08}.\footnote{However, some of these analyses have
  detected evidence of galaxy assembly bias at very low (or sometimes
  unspecified/undetermined) significance \citep[e.g.,][]{thomas05,
    bernardi06, clemens06}.}  These studies utilized a variety of data
sets from high--resolution spectroscopic observations of relatively
small numbers of cluster members to composite spectra constructed from
early SDSS data.

In contrast, more recent analyses have yielded contradictory
conclusions. While \citet{thomas05} found cluster ellipticals to have
older stellar populations than their field counterparts at the same
velocity dispersion or stellar mass, subsequent analysis by
\citet{thomas07} concluded that there is no significant environmental
dependence to the slope or zero--point of the age--mass relation for
the ``bulk'' of the nearby elliptical population (e.g., at ages of
$\gtrsim 4$ Gyr). Instead, \citet{thomas07} claim that the trend
detected by \citet{thomas05} is driven entirely by a population of
``rejuvenated'' galaxies (i.e., ellipticals with recent
star--formation activity and thus ages $\lesssim \! 3$ Gyr), which
preferentially reside in low--density environs.

Our results disagree, in part, with the more--recent conclusions of
\citet{thomas07}. In agreement with their work, we find evidence for a
variation in the fraction of ``rejunvenated'' galaxies with
environment (that is, we find a residual age--density relation at
young stellar ages). However, as shown in Figure \ref{resid_fig} and
Figure \ref{select_fig}, even when restricting our sample to galaxies
with ages of $\log_{10}(t/{\rm yr}) > 9.5$ (i.e., excluding the
``rejuvenated'' population), we detect a strong relationship between
age and residual environment --- in conflict with the conclusions of
\citet{thomas07}. That is, we find that the correlation between
luminosity--weighted mean stellar age and environment at fixed mass
spans the entire range of ages probed by our various galaxy samples,
rather than being an effect driven by only those galaxies with young
stellar populations.

\begin{figure*}[tb]
\centering
\plotone{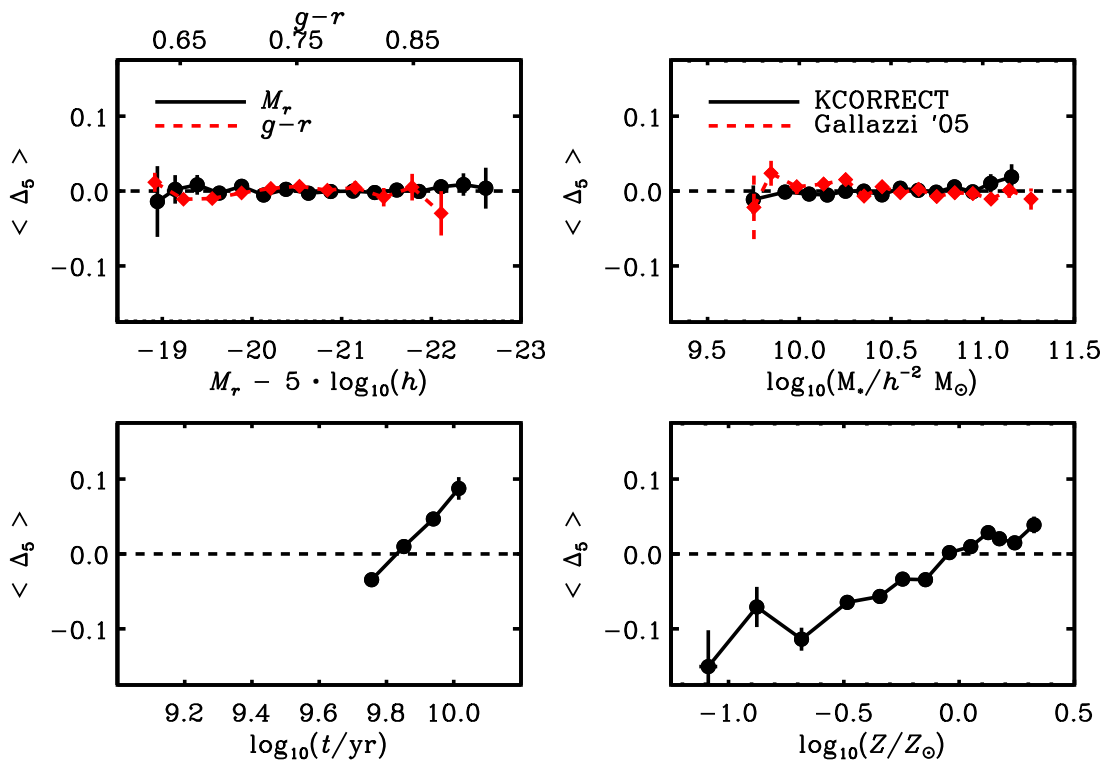}
\caption{The dependence of mean residual environment, $< \! \Delta_5
  \! >$, on absolute magnitude, rest--frame color, stellar mass,
  stellar age, and stellar metallicity for a sample of red--sequence
  galaxies with ages of $> \! 10^{9.6}$ yr and $r < 17.5$. The points
  and error bars correspond to the mean environments and $1\sigma$
  uncertainties in the means computed in distinct bins of each galaxy
  property. While we find no significant residual trend between
  environment and color, luminosity, or stellar mass, we observe a
  strong dependence of residual environment on stellar age and
  metallicity.}
\vspace*{0.1in}
\label{ages_fig}

\end{figure*}

To emphasize and further illustrate this point, we select multiple
subsamples according to various age selections (e.g., $t > 10^{9.5}$,
$10^{9.6}$, $10^{9.7}$, and $10^{9.75}$ yr). For each of these samples
of older galaxies, we independently repeat our analysis --- that is,
we measure and remove the color--luminosity--density relation in each
subsample and then examine the residual relationships between
environment and color, luminosity, stellar mass, age, and
metallicity. As shown in Figure \ref{ages_fig} for a sample of red
galaxies selected to have $\log_{10}(t/{\rm yr}) > 9.6$ and $r <
17.5$, we find no significant correlation between residual
environment, $\Delta_5$, and stellar mass. However, we do find
significant residual age--density and metallicity--density relations,
similar to those previously observed for samples selected according to
apparent brightness and morphology yet spanning the entire range of
ages probed (cf.\ Figure \ref{select_fig}).

The results as presented in Figure \ref{ages_fig} are independent of
the age limit employed to select the galaxy subsample; for samples
with ages $> \! 10^{9.5}$, $10^{9.7}$, and $10^{9.75}$ yr, we find no
significant variation in the residual relationship with environment
for stellar mass, age, and metallicity. Moreover, we find similar
results for subsamples of galaxies selected according to a
mass--dependent age criterion, analogous to that employed by
\citet{thomas07}. When removing the set of galaxies referred to as
``rejuvenated'' by \citet{thomas07} (i.e., those with younger stellar
populations), we still find a strong age--density relation at fixed
stellar mass within the population of galaxies with older stellar
populations (i.e., the ``bulk'' population according to
\citealt{thomas07}).

Finally, when defining subsamples selected on morphology and apparent
brightness as well as age (e.g., $r < 17$, $n > 2.5$, and
$\log_{10}(t/{\rm yr}) > 9.7$), we find trends between residual
environment and stellar age and metallicity analogous to those
presented in Figure \ref{select_fig}. However, by excluding those
galaxies with younger stellar populations in combination with
morphological and apparent magnitude selections, the number of sources
at low metallicities (e.g., $\log_{10}(Z/Z_{\sun}) < -0.5$) is
significantly diminished such that the range in metallicities probed
is limited and the resulting residual metallicity--density relation is
only statistically significant within the metal--rich regime (i.e.,
$\log_{10}(Z/Z_{\sun}) \gtrsim -0.35$). Still, altogether, this
analysis illustrates that our results are inconsistent with the
conclusions of \citet{thomas07}, such that we find a relationship
between stellar age and environment at fixed mass that spans all ages
probed in our study (i.e., the age--density trend at fixed mass is not
solely driven by galaxies with young stellar populations
preferentially residing in underdense environs). Additional evidence
for this conclusion, based on independent statistical tests, is
presented in the Appendix.

Other recent analyses have, like \citet{thomas07}, found little
evidence of galaxy assembly bias at the massive end of the red
sequence based on archaeological investigations of local
galaxies. Studying a dozen early--type galaxies in the Coma cluster in
detail, \citet{trager08} also find no difference between the inferred
stellar ages of red--sequence galaxies in the cluster environment and
similar galaxies in the field. As discussed by \citet{trager08}, the
relatively young ages of the red galaxies in Coma are most likely
attributed to recent star--formation episodes. However, nearly all of
the galaxies in their sample show mean ages of $5$--$7$ Gyr, with none
having stellar populations as old as $\sim \! 10$ Gyr. Past
spectroscopic observations of early--type systems in Coma have
generally arrived at a broader range of stellar ages, with a
non--negligible population of galaxies with ages of roughly $10$ Gyr
\citep{moore01, poggianti01, moore02, nelan05, sanchez06}. However,
the \citet{trager08} data are of somewhat higher resolution and/or
signal--to--noise ratio, facilitating improved emission corrections
and precision in the resulting age estimates.

Accepting the accuracy of the \citet{trager08} age measurements, our
results are reconcilable if either the \citet{trager08} sample is a
biased tracer of the Coma early--type population (e.g., due to shot
noise in the selection of their small sample) or if there are no (or
at least very few) old early--type systems in the Coma cluster. If the
latter is true, then Coma would be unusual amongst massive clusters in
the local Universe. Coma is not a cool--core or cooling--flow cluster;
instead it has a cooling time $\gtrsim 10$ Gyr
\citep{kaastra04}. Thus, recent star formation at a significant level
within the central region of the cluster seems unlikely, especially
across the entire early--type population. However, Coma is not a
relaxed system; instead, it contains significant substructure
\citep[e.g.,][]{biviano96, neumann03} and is observed to be undergoing
a merger/accretion event \citep[e.g.,][]{mellier88, neumann01}. As
highlighted by \citet{gerhard07}, a large fraction $(\sim \! 30\%)$ of
galaxies in Coma are likely associated with the accretion of the NGC
4839 subcluster. If Coma is a currently--forming supercluster, then
perhaps recent episodes of star formation, associated with
accretion/merger events, can explain the relatively young ages
measured for the \citet{trager08} sample of early--type galaxies.

All told, the young ages of early--type galaxies in Coma as found by
\citet{trager08} persists as a significant outlier relative to the
results of other studies of local clusters, including other
contemporary work studying Coma. For example, recent results from
\citet{smith09}, employing spectroscopic observations of comparable
quality to those of \citet{trager08} and spanning a larger sample of
early--type galaxies extending to lower masses, find a significant
population of galaxies with old stellar populations ($t \sim 10$ Gyr)
in the core of Coma. Moreover, \citet{caldwell03} find massive
early--type systems in Virgo with similarly old luminosity--weighted
ages ($t > 9$ Gyr). While the sample of \citet{smith09} is biased to
lower masses than that of \citet{trager08}, one would expect this to
lead to smaller, not larger, ages than for the \citet{trager08}
sample, as more massive systems tend to contain older stellar
populations \citep[e.g.,][]{trager00a, trager00b, nelan05, rogers08,
  michielsen08, smith09, matkovic09}.

There are several other recent studies of local field and cluster
galaxies that have found only weak statistical evidence for a
relationship between stellar age and environment at fixed mass. For
example, \citet{gallazzi06} studied the environmental dependence of
the scatter in the age--stellar mass relationship using the same
catalog of stellar ages and metallicities as employed in this
work. They find a relatively weak correlation between environment and
age such that galaxies outside of the most dense environments tend to
have younger stellar populations. The analysis of \citet{gallazzi06},
however, was limited in its scope due to a rather small sample size
($< \! 2000$ galaxies) for which environment estimates were available
from \citet{kauffmann04}. Our analysis expands upon this earlier work
by markedly increasing the sample size with environment information
and by utilizing a more sensitive statistical methodology.

While \citet{clemens06} and \citet{bernardi06} both conclude that
there are differences between the ages of massive ellipticals in
differing environments locally, with systems in high--density regions
being roughly $1$ Gyr older at fixed velocity dispersion, evidence for
a significant age--density relation at fixed mass in their samples is
considerably weaker than that presented herein and more in line with
the results of \citet{thomas07}. For example, \citet{clemens06} find a
significant difference between the ages of ellipticals in high-- and
low--density environs over only a limited range in velocity dispersion
(or specifically for only 2 of their 7 discrete bins in velocity
dispersion), with the differences (in those two bins) significant at
less than $2\sigma$ \citep[see Figure 10 of][]{clemens06}. Similarly,
in the work by \citet{bernardi06}, the apparent difference in average
age between ellipticals in high--density and low--density environs
depends on the particular Balmer absorption line (H$\beta$ versus
H$\gamma_{\rm F}$) employed to constrain stellar age, with ages based
on H$\beta$ showing no significant difference as a function of
environment and with H$\gamma_{\rm F}$ yielding age variations
significant at the approximately $2\sigma$ level \citep[see Table 9
of][]{bernardi06}.

The lower significance at which \citet{bernardi06} and
\citet{clemens06} find a trend between stellar age and environment at
fixed mass likely results from the smaller sample sizes, alternate
methods of measuring environment, and different statistical methods
employed in the analysis. For example, by selecting samples drawn from
the extremes of their measured environment distributions, both
\citet{bernardi06} and \citet{clemens06} are limited to final sample
sizes of $< \! 9000$ and $< \! 4000$ galaxies, respectively. On top of
that, \citet{clemens06} further reduce the statistical power of their
sample by dividing into course environment bins and then subdividing
into bins of velocity dispersion \citep[e.g., see Figure 10
of][]{clemens06}. While \citet{clemens06} uses an environment measure
similar to ours, \citet{bernardi06} use the 3--dimensional distance to
the $10^{\rm th}$--nearest neighbor with $M_r < -21.5$ as an
environment estimator. By restricting the tracer population to
relatively luminous galaxies, while also using a $10^{\rm th}$--
versus $5^{\rm th}$--nearest--neighbor distance (as used in this
work), \citet{bernardi06} are tracing environment on much larger
scales, perhaps missing the dependence of galaxy properties on the
more local galaxy density. Moreover, measuring distances in
3--dimensions versus in projection leads to greater sensitivity to the
effects of redshift--space distortions, which smear out the galaxy
distribution in environments such as clusters
\citep{cooper05}. Finally, along with using the distance to the
$10^{\rm th}$--nearest neighbor, \citet{bernardi06} use the distance
to the nearest galaxy cluster as a means for identifying high--density
environments, thereby significantly biasing their sample against
group--like environs.  Altogether, these effects likely lead to the
marginally--significant trends between stellar age and environment at
fixed mass presented by \citet{bernardi06} and \citet{clemens06}. In
contrast, our methodology, which utilizes the entire galaxy sample
(versus only studying galaxies in extreme environs), yields a strong
age--density relation at fixed mass.

\subsubsection{Metallicity--Density Relation}

In addition to a strong residual age--density relation, we also find a
relatively weak residual metallicity--density relation, where more
metal--rich galaxies favor regions of higher galaxy density relative
to galaxies of like stellar mass. However, the significance of this
metallicity--density relation decreases as the sample is restricted to
more centrally--concentrated (i.e., more bulge--dominated) or less
elongated galaxies (cf.\ \S \ref{sec_disc1}), such that a weak (though
statistically significant) correlation is observed between mean
residual environment, $< \!  \Delta_5 \! >$, and stellar metallicity
for the bulk of the galaxy samples studied here.

Previous analyses of early--type galaxies within local clusters have
generally found no evidence for a metallicity--density relation at
fixed mass, in contrast to our results. For instance, studies of line
indices in nearby clusters \citep[e.g.,][]{guzman92, carter02} find a
significant dependence of the Mg$_2$--$\sigma$ relation on
cluster--centric distance, where early--type galaxies near the core
exhibit stronger Mg$_2$ absorption than their counterparts in the
outskirts of the cluster. As \citet{guzman92} only measure one
spectral index, they are unable to disentangle age and metallicity;
although they do conclude that a dependence of stellar age on cluster
radius might be responsible for the observed variations in Mg$_2$
absorption strength \citep[see also][]{bernardi98}. A similar
conclusion is reached by \citet{terlevich01} studying the $U-V$ color
of galaxies in the Coma cluster; color, however, is similarly unable
to distinguish variations in age from variations in metallicity.

Folding in additional spectral indices to separate metallicity and age
effects, \citet[][see also \citealt{kuntschner98}]{carter02} infer a
significant relationship between metallicity and radius in
Coma. However, this metallicity--density relation does not take into
account the relationship between mass (or luminosity) and radius,
minimizing its usefulness in a comparison to our analysis. More recent
work by \citet{smith06}, measuring a larger number of indices ($12$)
for a significantly larger sample of galaxies ($\sim \! 3000$) in $94$
nearby clusters, finds several significant relationships between
spectral indices and cluster--centric radius at a given velocity
dispersion. Using these indices to distinguish between age and
metallicity effects, they conclude that there is a significant age
gradient in local clusters, such that systems in cluster cores have
older stellar populations, while there is no significant metallicity
gradient detected \citep[see also][]{spolaor09}.

Other recent analyses spanning a broad range of environments in the
local Universe have also found little evidence for a significant
relationship between stellar metallicity and local galaxy density at
fixed mass. For example, both \citet[][see also
\citealt{clemens09}]{clemens06} and \citet{bernardi06}, find no
significant metallicity--density trend at fixed velocity dispersion
for nearby elliptical and lenticular systems \citep[see
also][]{gallazzi06}. Furthermore, while \citet{thomas05} conclude that
massive early--type galaxies in local clusters are, on average, $\sim
\! 0.05$--$0.1$ dex more metal--poor than their counterparts in the
field, the significance of this result is unsubstantiated
statistically, with no uncertainties given for the coefficients of the
linear fits to the mean $Z$--$\sigma$ relations in the high-- and
low--density galaxy samples \citep[see Equation 1
of][]{thomas05}. Considering the small size of the galaxy sample (only
$124$ galaxies in total) employed by \citet{thomas05}, the
uncertainties on the parameters of the linear fits is likely to exceed
a few percent, which would make the small measured differences in the
mean $Z$--$\sigma$ relations in the two environment regimes
statistically insignificant. Thus, \citet{thomas05}, like many other
authors, find no evidence for a metallicity--density relation within
the nearby early--type galaxy population.

Not only do we find a significant metallicity--density relation at
fixed mass for our most inclusive color--selected sample, but also for
our morphologically--selected subsamples. While this trend is weaker
than the age--density relation at fixed mass, it is significant
nonetheless. As discussed in the preceding section (\S
\ref{sec_degen}), a metallicity--density relation is unlikely to
result from correlated errors associated with the degeneracy between
constraints on stellar age and metallicity. Moreover, when comparing
apples to apples, the uncertainties in measured galaxy properties are
generally large enough that it is far easier to dilute an underlying
environment trend, such that it goes undetected, than it is to falsely
detect one. Thus, for many previous studies of local early--type
galaxies, it is likely that the same errors or limitations that
smeared out the age--density relation in the galaxy sample also
precipitated a non--detection of the metallicity--density
relation.

In contrast to our results and to the majority of results from the
aforementioned studies of local early--type galaxies, \citet[][see
also \citealt{rose94}]{kuntschner02} find that their sample of nearby
field ellipticals are more metal--rich (by $\sim \! 0.2$ dex) in
comparison to members of the Fornax cluster. This discrepancy between
the findings of \citet{kuntschner02} and those of our work (as well as
those of \citet{bernardi06}, \citet{clemens06}, among others) can
potentially be understood in terms of an age--metallicity
degeneracy. As discussed in more detail in \S \ref{sec_degen}, errors
in estimating stellar age are strongly correlated with errors in
metallicity, such that if a galaxy's age is overestimated, then its
metallicity will correspondingly be underestimated. Thus, the
metallicity--density relation reported by \citet{kuntschner02} could
be reconciled with other observations, if the \citet{kuntschner02}
age--density relation is correspondingly stronger than that found in
other analyses. The plausibility of this scenario is supported by the
larger difference in age between massive ellipticals in field and
cluster environments given by \citet[][$\sim \! 2$--$3$
Gyr]{kuntschner02} versus that found by \citet[][$\sim \! 1$
Gyr]{bernardi06}.\footnote{Note that Table 9 of \citet{bernardi06}
  lists an age difference between high-- and low--density environs
  that is more in line with $\sim \! 0.6$ Gyr.}

The analysis by \citet{kuntschner02} suffers even more from the
problem of small sample size ($< \! 30$ galaxies in total). In
addition, there was a strong difference in the morphological
composition of their field and cluster samples, with the field
population containing a larger percentage of S0s (versus
ellipticals). However, the \citet{kuntschner02} study compared the
ages of galaxies in different environments at fixed $B$--band absolute
magnitude and not at fixed mass (i.e., not at fixed stellar mass or
velocity dispersion). Thus, their field sample may have been
increasingly dominated by younger galaxies at lower masses, which
appeared bright in the $B$--band due to recent star--formation
activity. Given the observed correlation between mass and metallicity
(independent of environment), such a bias in the sample would work to
decrease the strength of the metallicity--density relation detected by
\citet{kuntschner02}.

Recognizing the correlation between stellar metallicity and age
observed for local massive early--type systems, where galaxies of a
given mass with older stellar populations tend be more metal--poor
\citep[e.g.,][]{gallazzi05, sanchez06}, it might be expected that an
age--density relation at fixed mass would naturally be accompanied by
some sort of metallicity--density relation, like that observed by
\citet{kuntschner02}. However, there is large scatter in the
age--metallicity relation, especially at lower masses. Furthermore,
stellar age and metallicity measure different aspects of a galaxy's
formation history; while age traces the time since the bulk of the
stars were formed (or in the case of luminosity--weighted mean stellar
age, the time since the last significant episode of star formation),
metallicity reflects the balance of feedback and accretion processes
in concert with the star--formation efficiency (that is, the
efficiency with which gas is turned into stars). For example, two
galaxies forming from pristine gas reservoirs and following identical
formation histories, except for an offset in time, will contain
stellar populations of comparable metallicity, but distinct ages. Our
results suggest that not only did red--sequence galaxies in
high--density environments form earlier than their counterparts in
low--density regions, but [1] feedback and/or accretion processes are
modulated by the local environment such that the high--density regime
favors greater metal abundances and/or [2] the typical star--formation
efficiency of galaxies in high--density regions or at higher redshift
exceeds that of galaxies in low--density environs or at lower
redshift.

Some evidence for the former scenario has been observed in a study of
star--forming galaxies at $z \sim 0.1$ by \citet{cooper08b}. Using the
gas--phase oxygen abundance measurements of \citet{tremonti04}, they
show that local environment is correlated with a non--negligible
portion of the scatter in the mass--metallicity relation, such that
galaxies in higher--density environs at a given mass tend to be more
metal--rich. As discussed in more detail by \citet{cooper08b}, there
are many physical scenarios that might lead to galaxies having a
higher metal abundance in high--density regions versus in low--density
regions. For example, processes such as ram--pressure stripping or
galaxy harassment \citep{gunn72, moore96}, which only operate in
extreme high--density environs, could remove the outer portion (i.e.,
the most metal--poor portion) of a galaxy's gaseous halo, thereby
inflating the relative metal abundance of gas reservoirs within
galaxies in rich groups and clusters. Similarly, outflows and winds
driven from galaxies in high--density environments might be less
capable of expelling enriched gas versus their counterparts in
underdense regions. 

There is also evidence to support the possible higher star--formation
efficiencies of star--forming galaxies at higher redshift relative to
local star-forming galaxies. For example, studies of sub--millimeter
galaxies \citep[SMGs,][and references therein]{chapman05} as well as
luminous and ultra--luminous infrared galaxies \citep[LIRGs and
ULIRGs,][]{aaronson84, sanders96} at $z > 1$ typically find higher
surface densities of star formation for a given surface density of
molecular gas \citep[][]{bouche07, daddi08, tacconi09}. Such a
correlation between star--formation efficiency and redshift in concert
with a galaxy assembly bias where galaxies in high--density environs
formed their stars earlier (i.e., at higher redshift) than their
counterparts in low--density environs would yield a corresponding
correlation between environment and star--formation efficiency such
that galaxies in overdense regions are more efficient at converting
gas into stars (i.e., have higher stellar metallicity).

\subsection{The Impact of Mergers}

The stellar populations of galaxies are not entirely formed via
in--situ star formation. Rather, much of a galaxy's stellar mass is
assembled via accretion events (i.e., mergers). In addition to
increasing the stellar mass of a galaxy, mergers can also have a
substantial impact on the age of the resulting system. While gas--rich
(or ``wet'') mergers will tend to induce bursts of star formation
thereby significantly decreasing the inferred luminosity--weighted
mean stellar age of the system, dissipationless (or ``dry'') merging
of two systems with older stellar populations will have a less severe
impact on the inferred age of the merger product, yielding a
passively--evolving descendant with a stellar age that is simply the
luminosity--weighted average of the antecedents' ages.

In addition, mergers tend to occur in regions of higher galaxy
density, such as galaxy groups \citep[e.g.,][]{cavaliere92,
  mcintosh08, hester09, ideue09, lin09}. Given this correlation with
environment, it is important to understand the potential impact of
hierarchical assembly on our results. Since mergers increase the
stellar mass of the system and since galaxies with smaller stellar
masses tend to have younger stellar populations
\citep[e.g.,][]{gallazzi05, graves09}, galaxies that have recently
undergone a merger are likely to be biased towards lower ages relative
to other galaxies of like stellar mass. In the case of gas--rich
mergers, this effect is further accentuated by the decrease in age
associated with the merger--induced burst of star formation. Thus,
mergers would tend to decrease the average age of galaxies at a given
stellar mass in the high--density regime, thereby decreasing the
amplitude of the observed age--density relation at fixed mass. In all,
the impact of mergers would be to smear out the residual age--density
relation shown in Figure \ref{resid_fig}.

\subsection{The Evolution of Post--Starburst or Post--Quenching Galaxies}

Our analysis of red galaxies in the local Universe shows a clear
relationship between stellar age and environment, where galaxies with
older stellar populations tend to reside in higher--density
environments relative to younger galaxies of like stellar mass. Such a
correlation between stellar age and environment could naturally
explain the evolution in the clustering of post--starburst (otherwise
known as K+A or post--quenching) galaxies at $z < 1$, as observed by
\citet{yan09}.

Using data drawn from the SDSS and from DEEP2, \citet{yan09} studied
the distribution of environments for K+A galaxies \citep{dressler83}
at $z \sim 0.1$ and at $z \sim 0.8$, relative to the red and blue
galaxy populations at those redshifts. They find that post--starburst
galaxies at low redshift have an environment distribution similar to
that of blue galaxies, favoring regions less dense than those
typically inhabited by red galaxies \citep[see also][]{hogg06};
however, at higher redshift, they find post--starburst systems favor
environments more similar to that of red galaxies.

If post--starburst (or K+A) galaxies have their star--formation
quenched in the same type of environment (e.g., a density level
corresponding to a galaxy group) at all redshifts less than $z \sim
1$, then this evolution in the environment distribution of
post--starbursts relative to that of the red galaxy population from $z
\sim 0.8$ to $z \sim 0.1$ can naturally be explained in terms of
galaxy assembly bias. On average, a red galaxy at low redshift will
have had a longer amount of time since having its star formation
quenched; thus, the environment of a low--$z$ red galaxy will have had
more time to increase in density relative to that in which the
quenching occurred (i.e., the environment in which the K+A phase of
evolution occurs).  In this scenario, the time since quenching (or
time since assembly --- thus, stellar age) for a galaxy on the red
sequence is naturally correlated with the local galaxy density or
environment. Our results support this potential explanation for the
evolution of the environment distribution of post--starburst galaxies
at $z < 1$. The observed correlation between stellar age and
environment at fixed stellar mass supports (although does not prove)
the picture in which K+A galaxies are found in the same type of
environment at $z < 1$.

\subsection{Implications for the Environmental Dependence of the Type
  Ia SN Rate} 

The typical age of a stellar population has been shown to be strongly
connected to the type Ia supernova (SN) rate, with young star--forming
galaxies having much higher SN Ia rates than quiescent galaxies with
older stellar populations \citep{sullivan06}. The increase in the SN
Ia rate with star--formation rate is commonly attributed to the
existence of two types of Ia events: ``prompt'' SNe Ia, which likely
result from the evolution of somewhat massive stars and are thus
correlated with star formation, and ``delayed'' SNe Ia, which are
thought to be the evolutionary outcome of less massive stars and
therefore connected to the underlying stellar mass of the galaxy.

Within quiescent systems, the delayed component of the Ia rate
dominates. However, even for systems of like mass and lacking ongoing
star formation, the SN Ia rate is still expected to depend on the age
of the stellar population; some theoretical predictions for both
single--degenerate and double--degenerate progenitor scenarios suggest
that the SN Ia rate for a single--aged stellar population should
increase with time, reaching a peak around an age of $\sim \! 10^{8}$
yr, with a sharp decline towards later times \citep{greggio83,
  yungelson94}. A more recent analysis by \citet{greggio05} slightly
revises this general picture, finding a sharp reduction in the type Ia
rate for both single--degenerate and double--degenerate progenitors
beyond an age of $\sim \! 1$ Gyr.

Taking as given this theoretical connection between stellar age and
type Ia rate, our results regarding the correlation between age and
environment at fixed stellar mass along the red sequence suggest that
the type Ia rate in early--type galaxies should vary with
environment. In particular, our study indicates that the SN Ia rate
within quiescent early--type galaxies should be lower in high--density
environs relative to that found in low--density environs. The bulk of
our samples have typical luminosity--weighted mean stellar ages
greater than $1$ Gyr, placing these populations beyond the expected
peak in the type Ia rate and into a regime where the rate should be in
sharp decline.

Currently, the evidence is mixed regarding a potential correlation
between the SN Ia rate (per unit mass) and environment in the local
Universe. Most notably, \citet{mannucci08} concluded that the type Ia
rate is more than 3 times higher in cluster ellipticals relative to
field ellipticals, using a sample of 11 cluster and 5 field type Ia
events at $z < 0.04$, which would run counter to the theoretical
trend. However, the uncertainty of these environment--dependent rates
is quite large. In contrast, a recent analysis of a larger sample of
type Ia supernovae drawn from the SDSS--II Supernova Survey by
\citet{cooper09} found no significant relationship between local
galaxy environment and the likelihood of hosting a type Ia event
within the red galaxy population. Overall, such studies of the
environments of type Ia supernovae are currently limited by small
numbers. Results from future surveys \citep[e.g.,][]{sharon07, sand08}
will hopefully help in answering such outstanding questions.

\section{Summary}
\label{sec_summary}

Using the measurements of luminosity--weighted mean stellar age and
stellar metallicity from \citet{gallazzi05} and the local galaxy
overdensity estimates of \citet{cooper09}, we study the relationship
between galaxy properties and environment for various samples of
galaxies on the red sequence. In contrast to previous studies in the
local Universe, we do not solely focus on the early--type galaxy
population or on galaxies selected to reside in particular
environments (e.g., specific clusters such as Virgo or Coma); instead,
we define various galaxy samples drawn from the SDSS, controlling for
correlations between environment and color, morphology, mass, etc.,
while probing a broad and continuous range of environments. Our
principal results are as follows:

\begin{enumerate}

\item After removing the mean color--luminosity--environment relation
  on the red sequence, we find a strong residual relationship between
  environment and stellar age (cf.\ Figure \ref{resid_fig}), such that
  galaxies with older stellar populations favor regions of higher
  galaxy overdensity relative to galaxies of like color and luminosity
  (i.e., like stellar mass).

\item When using subsamples restricted according to morphological,
  brightness, and color criteria, we find this residual age--density
  relation persists with no significant change in strength (cf.\
  Figure \ref{select_fig} and Table \ref{tab2}). 

\item Using stellar velocity dispersion to trace mass (in lieu of
  stellar mass), we find that the age--density relation at fixed mass
  persists (cf.\ Figure \ref{vdisp_fig}) and is even more statistically
  significant (cf.\ Table \ref{tab1} and Table \ref{tab3}).

\item We find a significant residual correlation between environment
  and stellar metallicity at fixed color and luminosity (cf.\ Figure
  \ref{resid_fig}). While a strong residual $Z$--density relation is
  observed for our most inclusive sample of red--sequence galaxies,
  enforcing selection criteria based on concentration, axis ratio,
  and/or brightness yields samples that exhibit weaker (though still
  statistically significant) residual metallicity--density relations
  (cf.\ Figure \ref{select_fig} and Table \ref{tab2}).

\item Our results are in conflict with the general conclusions of
  recent work by \citet{thomas07} and \citet{trager08}, which both
  indicate a lack of correlation between age and environment at fixed
  mass. While the results of \citet{trager08} may be due to the
  peculiarities of the Coma cluster (relative to other local clusters)
  and/or sample selection effects, the \citet{thomas07} findings are
  clearly not supported by our analyses. While we find evidence for a
  variation in the fraction of ``rejuvenated'' (or young) galaxies
  with environment \citep[in agreement with the results
  of][]{thomas07}, we also find a correlation between age and
  environment at fixed mass even for the ``bulk'' (or old) early-type
  galaxy population --- counter to the claims of \citet{thomas07}.

\item In agreement with several previous studies of early--type
  systems in the local Universe and with a multitude of studies at
  intermediate redshift, we find significant evidence for assembly
  bias on the red sequence, such that red, early--type galaxies that
  formed earlier in the history of the Universe are more strongly
  clustered today than their counterparts (of equal mass) that formed
  later.

\end{enumerate}

As discussed in \S \ref{sec_intro}, evidence for an assembly bias in
the galaxy formation process has commonly been explored in two
distinct manners: an evolutionary approach, which studies galaxies
over a range of redshift and directly infers evolution in the galaxy
properties, and an archaeological or paleontological approach, which
studies the stellar populations of nearby galaxies in detail with the
aim of deriving the evolutionary history from the current fossil
record. As evidenced by the work presented herein, the results from
these two methods are now arriving at a relatively consistent picture.
Studies of galaxies at $z < 1$ have shown that the population of red
(or early--type) galaxies was preferentially built--up in overdense
environments, such that red--sequence galaxies in dense environments
have generally assembled their stellar mass and developed elliptical
morphologies earlier than those in less dense regions
\citep[e.g.,][]{treu05, bundy06, cooper07, gerke07, capak07}. In
addition, studies of galaxy populations at yet higher redshifts ($z
\sim 2$--$3$) have detected populations of massive galaxies with
passively--evolving stellar populations \citep[e.g.,][]{labbe05,
  longhetti05, papovich06, stutz08} that are very strongly clustered
\citep[e.g.,][]{daddi03, quadri07, quadri08}.

Our results, which apply the aforementioned archaeological method,
show a general agreement with these results from studies at higher
redshifts, supporting a picture of assembly bias in the local galaxy
population. We find that at a given mass on the red sequence galaxies
with older stellar populations are found in more overdense
environments (i.e., are more strongly clustered) than their
counterparts with younger stellar populations. In contrast to some
previous studies, we find that this correlation between age and
environment at fixed mass is \emph{not} solely driven by a minority
population of galaxies that have experienced recent episodes of star
formation. Instead, we find a strong residual age--density relation
even among those red--sequence galaxies dominated by older stellar
populations (i.e., with ages $\gtrsim \! 5$ Gyr).

In the future, we look to quantify the variation in the age of stellar
populations as a function of environment, with the goal of
constraining theoretical models of galaxy formation and evolution. As
shown in the Appendix, preliminary analysis of only those galaxies
with older stellar populations (i.e., $\log_{10}(t/{\rm yr}) > 9.75$)
clearly demonstrates that there is a difference of \emph{at least}
0.25 Gyr between those systems residing in high--density environs and
those in low--density regions. Some recent analyses at intermediate
redshift \citep[e.g.,][]{vandokkum03, vdw05, vandokkum07} suggest that
while early--type galaxies in clusters are older than their field
counterparts, the difference in stellar age is less than would be
expected from current generations of galaxy formation models. More
quantitative analysis at low and intermediate redshift will be
required to definitively measure the relationship between stellar
population parameters and environment at $z < 1$ and thereby further
constrain the theoretical models.

\vspace*{0.25in} 

\acknowledgments Support for this work was provided by NASA through
the Spitzer Space Telescope Fellowship Program. A.G.\ is grateful for
support from the DFG's Emmy Noether Programme of the Deutsche
Forschungsgemeinschaft. M.C.C.\ would like to thank Michael Blanton
and David Hogg for their assistance in utilizing the NYU--VAGC data
products. This work benefited greatly from conversations with Romeel
Dav\'{e}, Daniel Eisenstein, and Dennis Zaritsky. Finally, we thank
the anonymous referee for their insightful comments and suggestions
for improving this work.

Funding for the SDSS has been provided by the Alfred P.\ Sloan Foundation,
the Participating Institutions, the National Science Foundation, the
U.S.\ Department of Energy, the National Aeronautics and Space
Administration, the Japanese Monbukagakusho, the Max Planck Society, and
the Higher Education Funding Council for England. The SDSS Web Site is
http://www.sdss.org/.

The SDSS is managed by the Astrophysical Research Consortium for the
Participating Institutions. The Participating Institutions are the American
Museum of Natural History, Astrophysical Institute Potsdam, University of
Basel, University of Cambridge, Case Western Reserve University, University
of Chicago, Drexel University, Fermilab, the Institute for Advanced Study,
the Japan Participation Group, Johns Hopkins University, the Joint
Institute for Nuclear Astrophysics, the Kavli Institute for Particle
Astrophysics and Cosmology, the Korean Scientist Group, the Chinese Academy
of Sciences (LAMOST), Los Alamos National Laboratory, the
Max--Planck--Institute for Astronomy (MPIA), the Max--Planck--Institute for
Astrophysics (MPA), New Mexico State University, Ohio State University,
University of Pittsburgh, University of Portsmouth, Princeton University,
the United States Naval Observatory, and the University of Washington.


\appendix

In this Appendix, we investigate further the impact of galaxies with
relatively young stellar populations on the measured relationship
between age and environment at fixed stellar mass. As discussed in \S
\ref{sec_disc}, \citet{thomas07} conclude that the observed dependence
of mean stellar age on environment at fixed stellar mass or velocity
dispersion within the local early--type population is entirely driven
by ``rejuvenated'' galaxies, that is, early--type galaxies with
relatively young stellar populations \citep[$\log_{10}(t/{\rm yr})
\lesssim 9.6$ from Figure 1 of][]{thomas07}. For the ``bulk'' of the
galaxy population (i.e., the quiescent early--types with older stellar
populations), \citet{thomas07} claim no correlation between age and
environment at fixed mass.

To determine if there is an age--density relation at fixed mass within
the ``bulk'' population, we define a subsample of red--sequence
galaxies at $r < 17$, with early--type morphologies ($n_r > 2.5$), and
with stellar ages above the limit $\log_{10}(t/{\rm yr}) > 9.6$, so as
to exclude the ``rejuvenated'' population. From this parent
population, we then select those galaxies with stellar masses of $9.8
< \log_{10}({\rm M}_{*}/h^{-2} {\rm M}_{\sun}) < 11$ (using the
KCORRECT mass estimates) and residing in high--density environs such
that $\log_{10}(1+\delta_5) > 1.2$. This sample of $3291$ galaxies
represents the extreme high--density tail of the environment
distribution (comprising $< \! 8\%$ of the $r < 17.5$ red--sequence
population within this mass and age regime). As a comparison sample,
we also select those galaxies within the same apparent magnitude,
morphology, stellar mass and age limits and with measured local
overdensities below the cosmic mean (i.e., $\log_{10}(1+\delta_5) <
0$), or the least overdense $\sim \! 40\%$ of the sample. Now, from
this relatively large set of galaxies in low--density environments, we
randomly draw galaxies so as to match the stellar masses and redshifts
of the high--density subsample. That is, for each galaxy in the
high--density subsample, we randomly draw a member of the low--density
subsample from within a 2--dimensional window of $\Delta
\log_{10}({\rm M}_{*}/h^{-2} {\rm M}_{\sun})^2 + \Delta z ^2 < 0.002$,
centered on the stellar mass and redshift of the particular galaxy in
the high--density subsample. Some galaxies are duplicated in this
low--density comparison sample; however, the relatively large size of
the parent low--density subsample ensures that duplication is minimal
such that $> \! 90\%$ of the comparison sample is unique, with no
individual galaxy included more than 3 times.

For these two samples, matched in stellar mass and redshift but drawn
from distinct portions of the environment distribution, we then
compare the distributions of stellar masses, ages, and
metallicities. As shown in Figure \ref{ages2_fig}, the stellar mass
distributions for the high--density and low--density samples are in
excellent agreement, as expected given the construction of the
populations. However, the distributions of stellar ages for the two
samples are quite distinct, with the low--density population skewed to
younger ages relative to their high--density counterparts. Similarly,
the galaxies in low--density regions tend to be more metal--poor.

Applying non--parametric statistical tests to the data support the
perceived differences in the binned age and metallicity distributions
shown in Figure \ref{ages2_fig}. Specifically, we use the one--sided
Wilcoxon--Mann--Whitney (WMW) U test \citep{mann47} and the two--sided
Kolmogorov--Smirnov (KS) test \citep{press92, wall03} in our
analysis. The result of each test is a $P$--value: the probability
that a value of the WMW U or KS statistic equal to the observed value
or more extreme would be obtained, if some ``null'' hypothesis
holds. Results with a $P$--value below $0.05$ (roughly corresponding
to $2\sigma$ for a Gaussian distribution) are considered to be
significant, while we term $P$--values less than $0.01$ as highly
significant. For more details regarding the WMW U and KS statistics,
refer to \citet{cooper09}.

Performing a one--sided WMW U test on the age distributions for our
high--density and low--density subsamples, we find that the galaxies
in low--density regions do have younger ages, with a $P$--value $\ll
0.01$. Thus, there is a much less than 1\% chance that we would
observe a difference this strong if both samples were drawn from the
same parent distribution. For the KS test, we arrive at a similar
result, also finding a $P$--value $\ll 0.01$ when comparing the age
distributions for the galaxies in overdense and underdense environs.
Thus, when limiting the sample under study to galaxies with older
stellar populations (i.e., the ``bulk'' population), we find that
red--sequence galaxies residing in low--density environs are younger
than galaxies of like stellar mass in high--density regions. Applying
these same tests to the metallicity distributions, we again find
highly significant results, with $P$--values $\ll 0.01$, confirming
that galaxies in low--density regions are metal--poor relative to
their counterparts with the same stellar mass in high--density
environments.

Repeating the analysis above, while increasing or decreasing the
stringency of the morphological--, $r$--band magnitude--, and
age--limits applied to the parent galaxy sample, we find that the WMW
U and KS tests yield similar results (cf.\ Table \ref{tab4}). For
example, when limiting the parent population to those galaxies with
increasingly older stellar populations (i.e., $r < 17$, $n_r > 2.5$,
and $t > 10^{9.75}$ yr), the distributions of ages for the
high--density and low--density samples are distinct at a
highly--significant level (again, $P \ll 0.01$). As shown in Figure
\ref{ages2_fig}, these $\log_{10}(t/{\rm yr}) > 9.75$ samples have age
distributions well fit by Gaussians, lacking any tail towards lower
stellar ages and thus sample the ``bulk'' population exclusively.

The results illustrated and enumerated in Figure \ref{ages2_fig} and
Table \ref{tab4}, which are inconsistent with the conclusions of
\citet{thomas07}, illustrate that there is a significant bias in the
assembly of local early--type galaxies with older stellar populations
such that galaxies in low--density regions formed later than their
counterparts in high--density environments. Computing the difference
between the mean and median of the age distributions for the low-- and
high--density subsamples with $t > 10^{9.75}$ yr, we find an offset of
roughly $0.25$ Gyr. This, however, is a lower limit to the age
difference between early--type galaxies in underdense and overdense
environments. Measurements of local environment are inherently noisy
(relative to other measures of galaxy properties like rest--frame
colors, luminosities, and even stellar masses), causing correlations
with the local galaxy density to be smeared out. Also, by using a
broad range in $\log_{10}(1+\delta_5)$ over which to select our
low--density subsample, we reduced the dynamic range of any intrinsic
age--environment correlation.

\setcounter{table}{3}

\begin{deluxetable*}{l  c  c c  c c  c c}
  \tablewidth{0pt} \tablecolumns{8}
  \tablecaption{\label{tab4} Results of Statistical Tests} 
\startdata \hline
  \multirow{2}{*}{Sample} & \multirow{2}{*}{$N_{\rm overdense}$} &
  \multicolumn{2}{c}{$\log_{10}({\rm M}_{*})$} &
  \multicolumn{2}{c}{$\log_{10}(t/{\rm yr})$} &
  \multicolumn{2}{c}{$\log_{10}(Z/Z_{\sun})$} \\
  & & $P_{\rm WMW}$ & $P_{\rm KS}$ & $P_{\rm WMW}$ & $P_{\rm KS}$ & $P_{\rm WMW}$ & $P_{\rm KS}$ \\
  \hline \hline
  $r < 17.5, t > 10^{9.6}$                     
  & 6016 & 0.40 & 0.87 & 0.00 & 0.000 & 0.000 & 0.000 \\
  $r < 17.5, t > 10^{9.75}$                     
  & 5266 & 0.37  & 0.84 & 0.00 & 0.000 & 0.000 & 0.000 \\
\hline
  $r < 17, n_r > 2.5, t > 10^{9.6}$          
  & 3291 & 0.47 & 0.95 & 0.00 & 0.000 & 0.000 & 0.000 \\
  $r < 17, n_r > 2.5, t > 10^{9.75}$ 
  & 3016 & 0.39  & 0.92  & 0.00 & 0.000 & 0.000 & 0.000 \\
\hline
  $r < 17, n_r > 2.5, b/a > 0.7, t > 10^{9.6}$          
  & 2016 & 0.40 & 0.92 & 0.00 & 0.00 & 0.000 & 0.000 \\
  $r < 17, n_r > 2.5, b/a > 0.7, t > 10^{9.75}$ 
  & 1901 & 0.46  & 0.96  & 0.00 & 0.00 & 0.000 & 0.000 \\
\hline

\enddata
\tablecomments{We tabulate the $P$--values, $P_{\rm WMW}$ and $P_{\rm
    KS}$, from comparing the stellar mass, age, and metallicity values
  for the low--density and high--density subsamples drawn from the
  listed parent samples, using the Wilcoxon--Mann--Whitney (WMW) U
  test and the two--sided Kolmogorov--Smirnov (KS) test. As discussed
  in the text, smaller values indicate a lower probability that the
  observed differences in the samples will occur by chance if they are
  selected from the same underlying parent distribution. Note that, as
  a one--sided test, the $P$--values for the WMW U test have a a
  maximum value of $0.5$. For each sample, the number of galaxies in
  the overdense subsample ($N_{\rm overdense}$) is listed as well.}
\end{deluxetable*}

\setcounter{figure}{7}
\begin{figure*}[tb]
\centering
\plotone{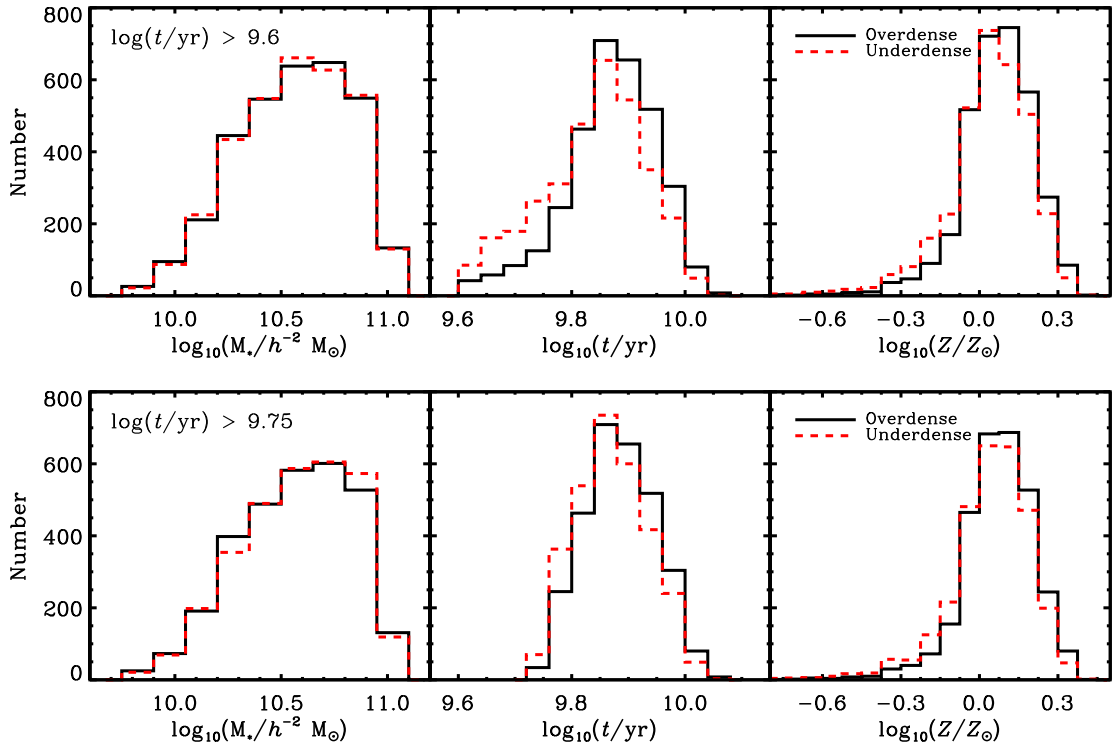}
\caption{The distributions of stellar masses (\emph{left}), ages
  (\emph{center}), and metallicities (\emph{right}) for red--sequence
  galaxies in high--density environments (solid black line) and for
  the comparison sample of galaxies in low--density regions, selected
  to have matching stellar masses and redshifts (red dashed line). The
  two environment--specific subsamples are selected from a parent
  population with $9.8 < \log_{10}({\rm M}_{*}) < 11$, $r < 17$, $n_r
  > 2.5$, and $\log_{10}(t/{\rm yr}) > 9.6$ (top) or $\log_{10}(t/{\rm
    yr}) > 9.75$ (bottom). Even when excluding those galaxies with
  young stellar populations, there remains strong evidence for
  assembly bias on the red sequence.}
\label{ages2_fig}
\end{figure*}

\end{document}